\pdfoutput=1 

\documentclass[11pt]{article}

\usepackage[square,numbers]{natbib}

\usepackage{geometry}
\usepackage{booktabs}
\usepackage{hyperref}
\usepackage{caption}
\usepackage{subcaption}

\usepackage{multirow}

\usepackage{xspace}
\usepackage{setspace}
\usepackage{enumerate}

\usepackage{censor}

\usepackage[T1]{fontenc} 
\usepackage{lmodern}     

\usepackage{listings}

\usepackage{pdflscape}

\usepackage{microtype}
\DisableLigatures{}

\usepackage{xcolor}
\definecolor{codegreen}{rgb}{0.25,0.5,0.35}
\definecolor{codegray}{rgb}{0.5,0.5,0.5}
\definecolor{codepurple}{rgb}{0.6,0,0}
\definecolor{backcolour}{rgb}{0.95,0.95,0.92}
\definecolor{colorstring}{rgb}{0.5,0,0.35}
\definecolor{rltred}{rgb}{0.5,0,0}
\definecolor{rltgreen}{rgb}{0,0.5,0}
\definecolor{rltblue}{rgb}{0,0,0.5}
\definecolor{DarkGreen}{rgb}{0.00,0.60,0.00}
\definecolor{ScarletRed}{rgb}{0.80,0.00,0.00}
\definecolor{blizzardblue}{rgb}{0.67, 0.9, 0.93}
\definecolor{green-yellow}{rgb}{0.68, 1.0, 0.18}
\definecolor{dkgreen}{rgb}{0,0.6,0}
\definecolor{gray}{rgb}{0.5,0.5,0.5}
\definecolor{mauve}{rgb}{0.58,0,0.82}
\definecolor{lightgrey}{rgb}{0.90,0.90,0.90}
\definecolor{grey}{gray}{0.75}
\definecolor{light-gray}{gray}{0.80}

\lstdefinestyle{mystyle}{
	backgroundcolor=\color{backcolour},   
	commentstyle=\color{codegreen},
	keywordstyle=\color{colorstring}\bfseries,
	numberstyle=\ttfamily\color{codegray},
	stringstyle=\color{codepurple},
            basicstyle={\scriptsize\ttfamily},
	breakatwhitespace=false,         
	breaklines=true,                 
	captionpos=b,                    
	keepspaces=true,                 
	numbers=left,                    
	numbersep=2pt,                  
	showspaces=false,                
	showstringspaces=false,
	showtabs=false,                  
	tabsize=2
}
\newcommand{\bBOXRT}{{\sc bBOXRT}\xspace}
\newcommand{\evo}{{\sc EvoMaster}\xspace}
\newcommand{\RESTest}{{\sc RESTest}\xspace}

\newcommand{\RestCT}{{\sc RestCT}\xspace}
\newcommand{\RESTler}{{\sc RESTler}\xspace}

\newcommand{\RestTestGen}{{\sc RestTestGen}\xspace}

\newcommand{\rpcncs}{{\emph{thrift-ncs}}\xspace}
\newcommand{\rpcscs}{{\emph{thrift-scs}}\xspace}
\newcommand{\csA}{{\emph{CS1}}\xspace}
\newcommand{\csB}{{\emph{CS2}}\xspace}
\newcommand{\csC}{{\emph{CS3}}\xspace}
\newcommand{\csD}{{\emph{CS4}}\xspace}

\newcommand{\rpcschemadto}{{\emph{RPCInterfaceSchemaDto}}\xspace}
\newcommand{\rpcfunctiondto}{{\emph{RPCActionDto}}\xspace}
\newcommand{\rpcparamdto}{{\emph{ParamDto}}\xspace}
\newcommand{\rpctypedto}{{\emph{TypeDto}}\xspace}
\newcommand{\rpcdatetype}{{\emph{RPCSupportedDataType}}\xspace}

\newcommand{\rpcresponse}{{\emph{ActionResponseDto}}\xspace}
\newcommand{\rpcexceptioninfodto}{{\emph{RPCExceptionInfoDto}}\xspace}
\newcommand{\rpcexceptiontype}{{\emph{RPCExceptionType}}\xspace}
\newcommand{\rpcexceptioncategory}{{\emph{RPCExceptionCategory}}\xspace}
\newcommand{\rpccustomizedresult}{{\emph{CustomizedCallResultCode}}\xspace}

\newcommand{\totalIndLoc}{1,444,045\xspace}
\newcommand{\avgIndLoc}{28,880.90\xspace}
\newcommand{\maxIndLoc}{89,303\xspace}
\newcommand{\minIndLoc}{2,019\xspace}
\newcommand{\maxIndLineCoverage}{59.16\%\xspace}
\newcommand{\avgIndLineCoverage}{17.49\%\xspace}
\newcommand{\totalIndfaults}{8,377\xspace}
\newcommand{\maxIndfaults}{2,250\xspace}
\newcommand{\avgIndfaults}{167.54\xspace}

\newcommand{\totalindcs}{50\xspace}
\newcommand{\totalcs}{54\xspace}
\newcommand{\totalLoc}{1,489,959\xspace}

\newcommand{\loc}{\#\textit{LoC}}
\newcommand{\interface}{\#\textit{Interfaces}\xspace}
\newcommand{\function}{\#\textit{Functions}\xspace}
\newcommand{\service}{\#\textit{Services}\xspace}
\newcommand{\class}{\#\textit{Classes}\xspace}
\newcommand{\tableNumber}{\#\textit{Tables}\xspace}
\newcommand{\serviceU}{\#\textit{U}\xspace}
\newcommand{\serviceD}{\#\textit{D}\xspace}

\usepackage{pgfplots} 

\definecolor{ForestGreen}{RGB}{34,139,34}
\definecolor{asparagus}{rgb}{0.53, 0.66, 0.42}

\usepackage{boxedminipage}
\newenvironment{result}%
{\smallskip
	\noindent
	\let\emph=\textbf
	\begin{boxedminipage}{\columnwidth}\begin{center}\em}%
		{\end{center}\end{boxedminipage}%
}

\newboolean{showcomments}
\setboolean{showcomments}{true} 

\ifthenelse{\boolean{showcomments}}{
	\newcommand{\nbc}[3]{
		{\colorbox{#3}{\bfseries\sffamily\scriptsize\textcolor{white}{#1}}}
		{\textcolor{#3}{\sf\small$\langle$\textit{#2}$\rangle$}}}
	
}{
	\newcommand{\nbc}[3]{}

}

\title{White-box Fuzzing RPC-based APIs with EvoMaster:\\ An Industrial Case Study}
\author{Man Zhang\textsuperscript{1}, Andrea Arcuri\textsuperscript{1}, Yonggang Li\textsuperscript{2}, Yang Liu\textsuperscript{2}, Kaiming Xue\textsuperscript{2}\\ \textsuperscript{1}Kristiania University College, Norway \\ \textsuperscript{2} Meituan, China}
\date{}

\begin{document}

\maketitle

\begin{abstract}
Remote Procedure Call (RPC) is a communication protocol to support client-server interactions among services over a network.
RPC is widely applied in industry for building large-scale distributed systems, such as Microservices.
Modern RPC frameworks include for example Thrift, gRPC, SOFARPC and Dubbo. 
Testing such systems using RPC communications is very challenging, due to the complexity of distributed systems and various RPC frameworks the system could employ.
To the best of our knowledge, there does not exist any tool or solution that could enable automated testing of modern RPC-based services.
To fill this gap, in this paper we propose the first approach in the literature, together with an open-source tool, for fuzzing modern RPC-based APIs.
The approach is in the context of white-box testing with search-based techniques.
To tackle schema extraction of various RPC  frameworks, we formulate RPC schema specification along with a parser that allows the extraction from source code of any JVM RPC-based APIs.
Then, with the extracted schema we employ search to produce tests by maximizing white-box heuristics and newly defined heuristics specific to RPC domain.
We built our approach as an extension to an open-source fuzzer (i.e., \evo), and the approach has been integrated into a real industrial pipeline that could be applied to real industrial development process for fuzzing RPC-based APIs.
To assess our novel approach, we conducted an empirical study with two artificial  and four industrial web services selected by our industrial partner.
In addition, to further demonstrate its effectiveness and application in industrial settings, 
we report results of employing our tool for fuzzing another \totalindcs industrial APIs autonomously conducted by our industrial partner in their testing processes.
Results show that our novel approach is capable of enabling automated test case generation for industrial RPC-based APIs (i.e., two artificial and \totalcs industrial).
We also compared with a simple grey-box technique and existing manually written tests. 
Our white-box solution achieves significant improvements on code coverage. 
Regarding fault detection, 
by conducting a careful review with our industrial partner of the tests generated by our novel approach in the selected four industrial APIs,   
a total of 41 real faults were identified, which have now been fixed.
Another \totalIndfaults detected faults are currently under investigation.

\end{abstract}



\section{Introduction}

It is a common practice in industry to develop large enterprise systems with microservice architectures~\cite{newman2015building,zhou2018fault,waseem2021design}.
For example, Meituan is a large e-commerce enterprise with more than 630 millions customers in China, with microservice systems like Meituan Select comprising more than 1000 different web services.
Testing this kind of systems is very complex, due to their distributed nature and access to external services such as databases.
There is a dire need in industry for automation for this kind of systems.

Although in the recent years there has been an interest in the research community on fuzzing REST web services~\cite{golmohammadi2022testing} (e.g., with tools like
Restler~\cite{restlerICSE2019},
RestTestGen~\cite{viglianisi2020resttestgen},
Restest~\cite{martinLopez2021Restest},
RestCT~\cite{wu2022icse},
bBOXRT~\cite{laranjeiro2021black},
and Schemathesis~\cite{hatfield2022deriving}),
to the best of our knowledge there is no work in the literature on the testing of modern Remote Procedure Call (RPC) web services.
There is a large body of knowledge in the scientific literature on the topic of software testing automation,
with several successful stories in many different software testing
domains~\cite{bertolino2007software,ali2009systematic,godefroid2020fuzzing}.
Addressing an important industrial testing problem for the first time does not start from scratch, especially when aiming at providing useful results for engineers in industry.
It rather builds on top of decades of scientific research on the topic.
On the one hand, some research challenges are similar to other domains (e.g., how to deal with SQL databases when fuzzing a web service~\cite{arcuri2020handling}, regardless of whether it is a  REST, GraphQL or RPC-based API).
On the other hand, scientific research and empirical evaluations are needed to address the specific peculiarities of each different software testing problem.
For example, to the best of our knowledge,
none of the existing fuzzers in the scientific literature can be directly applied on fuzzing RPC systems
without major engineering and scientific effort, as for example the API schemas and communication protocols are different.

As part of an industry-driven collaboration~\cite{garousi2019characterizing,arcuri2018experience,garousi2017,garousi2017industry,garousiindustry}, when we first tried to use our \evo fuzzer~\cite{arcuri2018evomaster} for RESTful APIs on the web services developed at Meituan, we could not apply it directly~\cite{meituanArxiv2022}.
We had to manually write REST APIs as wrappers for the RPC systems (which use Apache Thrift).
Not only it is time consuming, but also the generated tests are more difficult to use for debugging any found fault. 
Two web services were used as case study. 
Such study (with interviews and questionnaires among the developers at Meituan) pointed out to few research challenges, including the need for a native support for RPC systems for web service fuzzers. 
Such support not only requires not trivial engineering effort (our extension
to the existing fuzzer \evo required more than 10 000 lines of code, not including test cases), but also there are several scientific research challenges that need to be addressed to best handle  RPC-based APIs (as we will discuss in more details later in the paper).

In this paper, we provide a novel approach\footnote{
	\evo is open-source, and it is available at \url{www.evomaster.org}. 
	A replication package for this study is available at \url{https://github.com/anonymous-authorxyz/fuzzing-rpc}
}\label{foot:link} to automatically fuzz RPC-based APIs, built on top of \evo. 
To adapt to various RPC frameworks, in this approach, 
a RPC schema is defined to formulate the API specification that could document all necessary info to make a RPC call and possible responses (e.g., throwing exception, failure). 
The schema of the RPC-based services can be automatically extracted from the source code with our approach.
This allows to test the services developed with different RPC frameworks.
With the extracted schema, a test for a RPC-based API can be reformulated as an \textit{individual}, i.e., a sequence of RPC calls under a certain state of the API (e.g., database if it has).
Thus, search-based techniques (such as the MIO algorithm~\cite{arcuri2018test}) can be employed to evolve tests (e.g., seek various values of input parameters of RPC calls in order to cover more code and find more faults).
To better solve our testing problems with search, we define new heuristics specific to RPC domain. 
The approach was implemented as an extension of \evo and has been integrated into an industrial pipeline.
To assess effectiveness of our novel approach and its application on industrial context, we empirically compared it with a grey-box technique with two artificial and four industrial RPC-based APIs, and further reported its performances on 50 industrial APIs in real industrial settings.  
Main contributions of the paper include: 
\begin{enumerate}
	\item the first approach in the literature for fuzzing RPC-based APIs; 
	\item an open-source tool support (i.e., an extension to the existing fuzzer \evo);
	\item an empirical study carried out in industrial settings that involves in total \totalcs industrial RPC-based APIs comprising \totalLoc lines of codes (computed with JaCoCo) for business logic; 
	\item an in-depth analysis on 4 selected industrial APIs with our industrial partner; and
	\item identifying lessons learned and research challenges 	that must be addressed before better results can be obtained. 
\end{enumerate}

The paper is organized as follows.
Section~\ref{sec:background} provides the needed background information to better understand the rest of the paper.
Section~\ref{sec:related} analyzes related work.
The details of our novel approach are presented in Section~\ref{sec:fuzzing}.
Our empirical study is discussed in Section~\ref{sec:empirical}, followed by lessons learned in Section~\ref{sec:lessons}.
Threats to validity are discussed in Section~\ref{sec:threats}.
Finally, we conclude the paper in Section~\ref{sec:conclusion}. 
%

\section{Background}
\label{sec:background}

\subsection{Remote Procedure Call (RPC)}
\label{sub:rpcbg}

\begin{figure}[ht!]
	\centering
	\includegraphics[width=.8\linewidth]{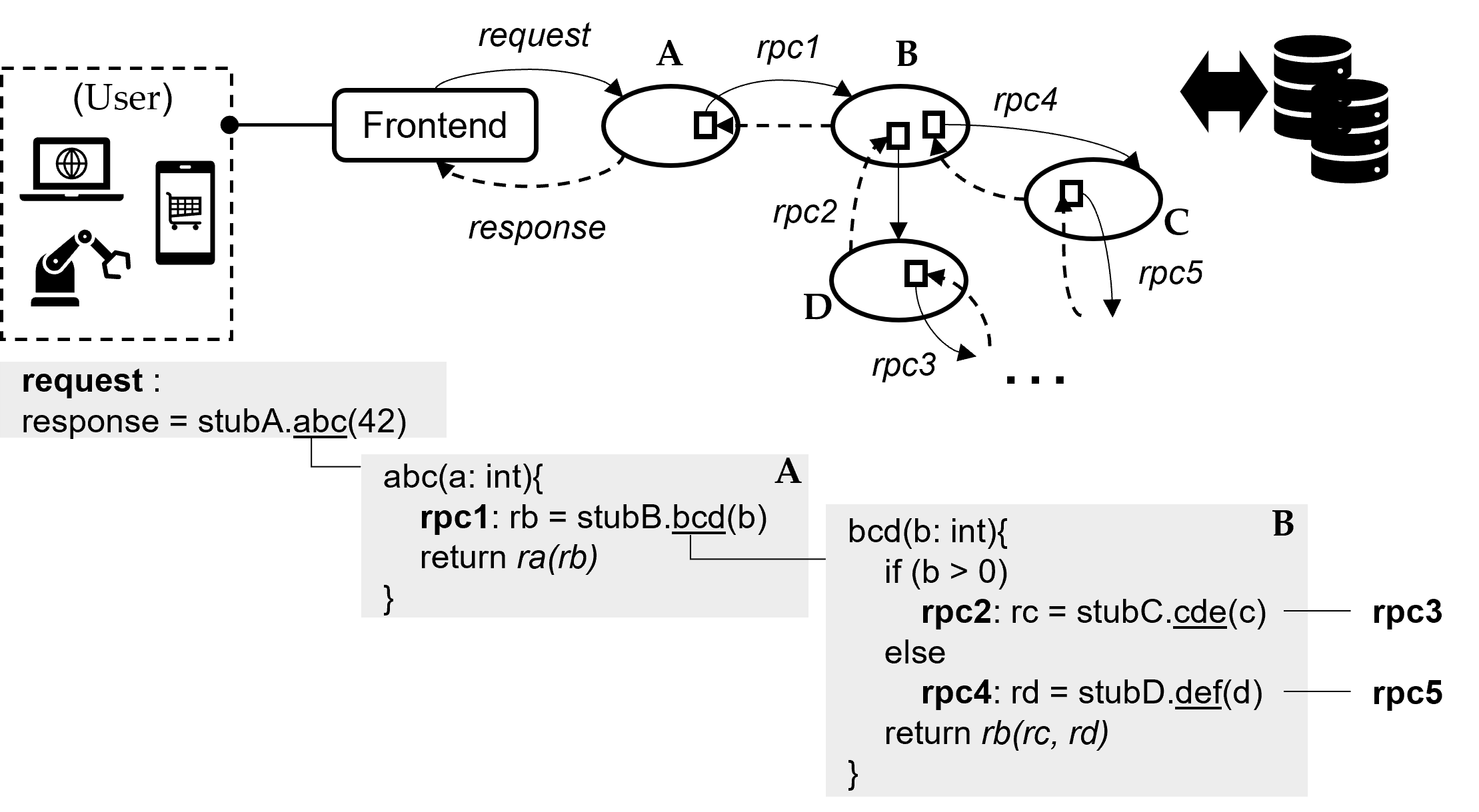}
	\caption{RPC-based APIs in Microservices}
	\label{fig:rpc-context}
\end{figure}

Remote Procedure Call (RPC) enables to call methods in other processes, possibly on a different machine, communicating over a network.
This is a common practice in distributed systems, particularly in microservice architectures~\cite{newman2015building}.  
There are different frameworks to develop RPC-based APIs,
like for example 
Apache Thrift~\cite{thrift} (originally from Facebook),
Apache Dubbo~\cite{dubbo} (originally from Alibaba),
gRPC~\cite{gRPC} (from Google) 
and SOFARPC~\cite{sofarpc} (from Alibaba).
All these popular frameworks were developed to address the scale of large distributed systems.
Compared to other types of web services (e.g., RESTful APIs), RPC-based APIs aim at optimizing performance at the cost of stronger coupling between client and server applications (there is no silver bullet). 
For example, given a schema for the API (e.g., a \texttt{.thrift} file for Thrift or a \texttt{.proto} file for gRPC), a compiler is used to create a server application (which then can be extended with the business logic of the API) and a client library.
A process that wants to communicate with the server API must include this client library, and use  these client-stubs to remotely call the API in the server process. 
For example (\textit{rpc1} in Figure~\ref{fig:rpc-context}), a client process (in the service \texttt{\texttt{A}}) would have a reference to a class-stub \texttt{B}, and, every time that \texttt{stubB.bcd()} is called, then the client-library will make a network call to execute \texttt{B.bcd()} on the server API.

The actual communications between the client and the server depend on the framework implementation, e.g., typically HTTP/2 using Protobuf for gRPC, but it can be configured to use other protocols.
Both Thrift and gRPC support the generation of client/server code in different languages (e.g., Java, C\# and JavaScript), whereas SOFARPC and Dubbo support only Java.

\begin{figure}
	\begin{subfigure}{1\textwidth}
		\begin{lstlisting}[language=java,basicstyle=\scriptsize,escapechar=©]
namespace java org.thrift.ncs

struct Dto {
	1: i32 resultAsInt,
	2: double resultAsDouble
}

service NcsService {

	Dto bessj(1:i32 n, 2:double x)
	...
}\end{lstlisting}
		\caption{Snippet of an example of RPC schema (i.e., \texttt{ncs.thrift}) specified with Thrift}
		\label{sub:thriftschema}
	\end{subfigure}\hfill

	\begin{subfigure}{1\textwidth}
		\begin{lstlisting}[language=java,basicstyle=\scriptsize,escapechar=©]
package org.thrift.ncs.client;

@SuppressWarnings({"cast", "rawtypes", "serial", "unchecked", "unused"})
@javax.annotation.Generated(value = "Autogenerated by Thrift Compiler (0.15.0)", date = "2021-10-21")
public class NcsService {

	public interface Iface {©\label{line:startthriftinterface}©

		public Dto bessj(int n, double x) throws org.apache.thrift.TException;©\label{line:endthriftinterface}©
		...
	}

	public static class Client extends org.apache.thrift.TServiceClient implements Iface {©\label{line:thriftclient}©

		public Dto bessj(int n, double x) throws org.apache.thrift.TException
		{
			send_bessj(n, x);
			return recv_bessj();
		}
		...
	}
}\end{lstlisting}
		\caption{Snippet of an example of an interface generated with Thrift framework}
		\label{sub:thriftjava}
	\end{subfigure}\hfill
	\begin{subfigure}{1\textwidth}
		\begin{lstlisting}[language=java,basicstyle=\scriptsize,escapechar=©]
@Test(timeout = 60000)
public void test() throws Exception {

	TTransport transport = new THttpClient("http://localhost:8080/ncs"); ©\label{line:url}©
	TProtocol protocol = new TBinaryProtocol(transport); ©\label{line:protocol}©
	NcsService.Client client = new NcsService.Client(protocol); ©\label{line:clientinstance}©

	org.thrift.ncs.client.Dto res_1 = null;
	{
		int arg0 = 577;
		double arg1 = 0.20491354575856158;
		res_1 = client.bessj(arg0,arg1); ©\label{line:rpcrequest}©
	}
	assertEquals(0, res_1.resultAsInt);  ©\label{line:startassert}©
	assertTrue(numbersMatch(0.0, res_1.resultAsDouble));©\label{line:endassert}©
}\end{lstlisting}
		\caption{An example of a Junit test for NcsService implemented with Thrift (see Figure~\ref{sub:thriftjava}
			)}
		\label{fig:testexample}
	\end{subfigure}

	\begin{spacing}{0.8}
		\raggedright \footnotesize * note that complete versions of the schema and its implementation with Thrift framework can be found at \url{https://github.com/EMResearch/EMB/tree/master/jdk_8_maven/cs/rpc/thrift/artificial/thrift-ncs}
	\end{spacing}

	\caption{An example of RPC schema, its automatically generated source code with Thrift framework and a test for the RPC-based service}
	\label{fig:thriftexample}
\end{figure}

\begin{figure}
	\begin{subfigure}{1\textwidth}
		\begin{lstlisting}[language=java,basicstyle=\scriptsize,escapechar=©]
syntax = "proto3";

option java_multiple_files = true;
option java_package = "org.grpc.ncs.generated";

service NcsService {

	rpc bessj(BessjRequest) returns (DtoResponse) {}
	...
}

message BessjRequest{
	int32 n = 1;
	double x = 2;
}

message DtoResponse {
	int32 resultAsInt = 1;
	double resultAsDouble = 2;
}\end{lstlisting}
		\caption{Snippet of an example of RPC schema (i.e., \texttt{ncs.proto}) specified with gRPC}
		\label{sub:gRPCschema}
	\end{subfigure}\hfill

	\begin{subfigure}{1\textwidth}
		\begin{lstlisting}[language=java,basicstyle=\scriptsize,escapechar=©]
package org.grpc.ncs.generated;

import static io.grpc.MethodDescriptor.generateFullMethodName;

@javax.annotation.Generated(
value = "by gRPC proto compiler (version 1.41.0)",
comments = "Source: ncs.proto")
@io.grpc.stub.annotations.GrpcGenerated
public final class NcsServiceGrpc {

	public static final String SERVICE_NAME = "NcsService";

	public static abstract class NcsServiceImplBase implements io.grpc.BindableService {©\label{line:startgRPCinterface}©

		public void bessj(org.grpc.ncs.generated.BessjRequest request,
		io.grpc.stub.StreamObserver<org.grpc.ncs.generated.DtoResponse> responseObserver) {
			io.grpc.stub.ServerCalls.asyncUnimplementedUnaryCall(getBessjMethod(), responseObserver);
		}©\label{line:endgRPCinterface}©
		...
	}

	public static final class NcsServiceBlockingStub extends io.grpc.stub.AbstractBlockingStub<NcsServiceBlockingStub> {©\label{line:gRPCclient}©

		public org.grpc.ncs.generated.DtoResponse bessj(org.grpc.ncs.generated.BessjRequest request) {
			return io.grpc.stub.ClientCalls.blockingUnaryCall(
			getChannel(), getBessjMethod(), getCallOptions(), request);
		}
		...
	}
	...
}\end{lstlisting}
		\caption{Snippet of an example of an interface generated with gRPC framework}
		\label{sub:gRPCjava}
	\end{subfigure}

	\begin{spacing}{0.8}
		\raggedright \footnotesize * note that complete versions of the schema and its implementation with gRPC framework can be found at \url{https://github.com/EMResearch/EMB/tree/master/jdk_8_maven/cs/rpc/grpc/artificial/grpc-ncs}
	\end{spacing}

	\caption{An example of RPC schema and its automatically generated source code with gRPC framework}
	\label{fig:gRPCexample}
\end{figure}

Figures~\ref{fig:thriftexample} and \ref{fig:gRPCexample} represent examples of a schema specified with different frameworks (i.e., Thrift and gRPC), and snippets of code of client-stub classes and server classes (e.g., interface/abstract classes) to implement/extend that are automatically generated by the framework based on the schema.
As shown in Figure~\ref{sub:thriftschema} for Thrift and Figure~\ref{sub:gRPCschema} for gRPC, the two schemas for \texttt{NcsService} have the same function, i.e., \texttt{bessj} that evaluates Bessel function by taking one \texttt{integer} and one \texttt{double} numbers as inputs then returning a Data Transfer Objects (\texttt{DTO}) which comprises same fields. 
Based on the schemas, the compiler could automatically generate source code of client libraries, such as \texttt{Client} class at line~\ref{line:thriftclient} in Figure~\ref{sub:thriftjava} and \texttt{NcsServiceBlockingStub} class at line~\ref{line:gRPCclient} in Figure~\ref{sub:gRPCjava}.
Then, with such client libraries, the functions of RPC-based API could be accessed.
For instance, snippet code shown in Figure~\ref{fig:testexample} represents an example of a test for the \texttt{NcsService} implemented with Thrift framework.
Lines~\ref{line:url} and \ref{line:clientinstance} in Figure~\ref{fig:testexample} represent how to instantiate a client to access the service, such as a URL with \textit{\url{http://localhost:8080/ncs}} and an accepted protocol to perform communications between client and service as \texttt{TBinaryProtocol}.
Line~\ref{line:rpcrequest} is to make a network call to \texttt{bessj} function with the instantiated client then receive a response, and lines~\ref{line:startassert} and \ref{line:endassert} show assertions on the response.

Besides the client-stub classes, the compiler also generate the source code which users could extend for implementing business logic of the services.
In this example, with Java language, Thrift outputs \texttt{Interface} class which comprises a list of methods to implement, and each method corresponds to a RPC function (see lines~\ref{line:startthriftinterface}--\ref{line:endthriftinterface} in Figure~\ref{sub:thriftjava}).
For gRPC, it is similar that the gRPC compiler outputs \texttt{abstract class} which comprises a list of methods to extend (see lines~\ref{line:startgRPCinterface}--\ref{line:endgRPCinterface} in Figure~\ref{sub:gRPCjava}).
As the examples, such methods in the classes define specifications about how to access the services.
In addition, a schema specification (e.g., \texttt{.thrift} file) might not be always available for a RPC-based API, and the service could be initially defined with programming language, such as SOFARPC\footnote{\url{https://www.sofastack.tech/en/projects/sofa-rpc/getting-started-with-sofa-boot/}} and Dubbo\footnote{https://github.com/apache/dubbo} with Java \texttt{interface} classes.
To fuzz RPC-based APIs, extracting specifications to access the API is a prerequisite.  
The various RPC frameworks allow  \textit{abstraction} classes (i.e., \texttt{Interface} and \texttt{abstract class} in Java) to define RPC-based APIs (we refer the classes which define RPC-based APIs as \textit{RPCInterface}s in later sections).
Thus, 
if we could enable an extraction of the specification based on such \textit{RPCInterface}s, 
it would generalize the application of the approach for fuzzing the APIs with different RPC frameworks (as we propose in this paper).

Moreover,
in the context of microservice architectures, as shown in Figure~\ref{fig:rpc-context}, the microservices comprise of a set of connected RPC-based APIs, and the API could have multiple stub-classes of other direct interacted APIs (e.g., \texttt{B} has stub-classes of the services \texttt{C} and \texttt{D}). 
Processing a request from the user typically involves multiple APIs.
With different inputs in the request, it could result in various sequences of RPC calls with different APIs.
For instance, assume that a user sends a request that results in a function call to \texttt{abc} of the service \texttt{A}. 
In order to provide a response to the user, it needs to involve multiple APIs (e.g., \texttt{B}, \texttt{C} and \texttt{D}) that could trigger various sequences of RPC communications, e.g., \textit{rpc1} $\rightarrow$ \textit{rpc2} $\rightarrow$ \textit{rpc3} or \textit{rpc1} $\rightarrow$ \textit{rpc4} $\rightarrow$ \textit{rpc5} as shown in Figure~\ref{fig:rpc-context}.
Note that this example illustrates communications among RPC-based services for processing one request.
To test one RPC-based API (e.g., \texttt{B} is the System Under Test (SUT)), \texttt{C} and \texttt{D} could be considered as external services of \texttt{B}, and the test for the API \texttt{B} would consist of a sequence of network calls to the SUT as the example shown in Figure~\ref{fig:testexample}.

\subsection{Automated System Testing}

In the scientific literature,
there has been a lot work on the automation of software testing~\cite{bertolino2007software}.
Different techniques have been investigated, like
search-based algorithms~\cite{ali2009systematic}
and symbolic execution~\cite{baldoni2018survey}.

System testing refers to the testing of applications as a whole, using the same input interfaces as the actual users.
The process of generating test cases to find errors in these systems (e.g., typically a crash) is often referred with the term \emph{fuzzing}~\cite{godefroid2020fuzzing}.

When dealing with the fuzzing of Web APIs, there are two main types of testing: \emph{black box} and \emph{white box} testing.
The difference is that in white box testing there is access to the internal details of the API, like its source code or bytecode.
This information can be exploited to design heuristics in order to improve the testing process, e.g., to increase metrics like code coverage.
 This can lead to higher fault detection, as a fault cannot be triggered if its code is never executed.
On the other hand, in black box testing, the API is treated as a black box, with no info on its internal details.

Black box testing is of more general application, as the API could be written in any programming language and running on a remote machine.
As white-box testing requires to analyze the source code, not only the API needs to be run locally, but also there are restrictions on the programming language it is implemented with.
Code analyses and white-box heuristics require complex engineering effort, and supporting several programming languages in the same tool might not be viable.
This also complicates scientific comparisons.
For example, it would be not viable to directly compare a fuzzer for C++ with one for Java.

Even in the cases in which the API is treated as a black-box, some general information is required to be able to fuzz it.
Typically, a specification (also known as \emph{schema}) is used to determine the type of inputs to send to the API.
Sending random bytes over a TCP connection would unlikely result in any meaningful message that the API will not discard immediately.
The type of schema will depend on the type of API, e.g., OpenAPI~\cite{Swagger} for RESTful APIs.
Alternatives are to use existing test cases or replayable traffic messages  as starting point, and then do small modifications to them to see how the API behaves.

Another common term used in the literature is \emph{grey box} testing.
This usually refers to a mix of black and white box testing, where only
partial information on the internal details of the API is available.
At times, this term is also used to specify when only
\emph{lightweight} code instrumentations are applied to the tested API.
A lightweight instrumentation could be just measuring code coverage, for example to drive the fuzzer to focus
on the least covered parts of the API.
A lightweight instrumentation is much easier to implement and start to experiment with compared to a full blown search-based or symbolic execution approach.

\subsection{EvoMaster}
\label{sub:evomaster}

\begin{figure}[ht!]
	\centering
	\includegraphics[width=0.5\linewidth]{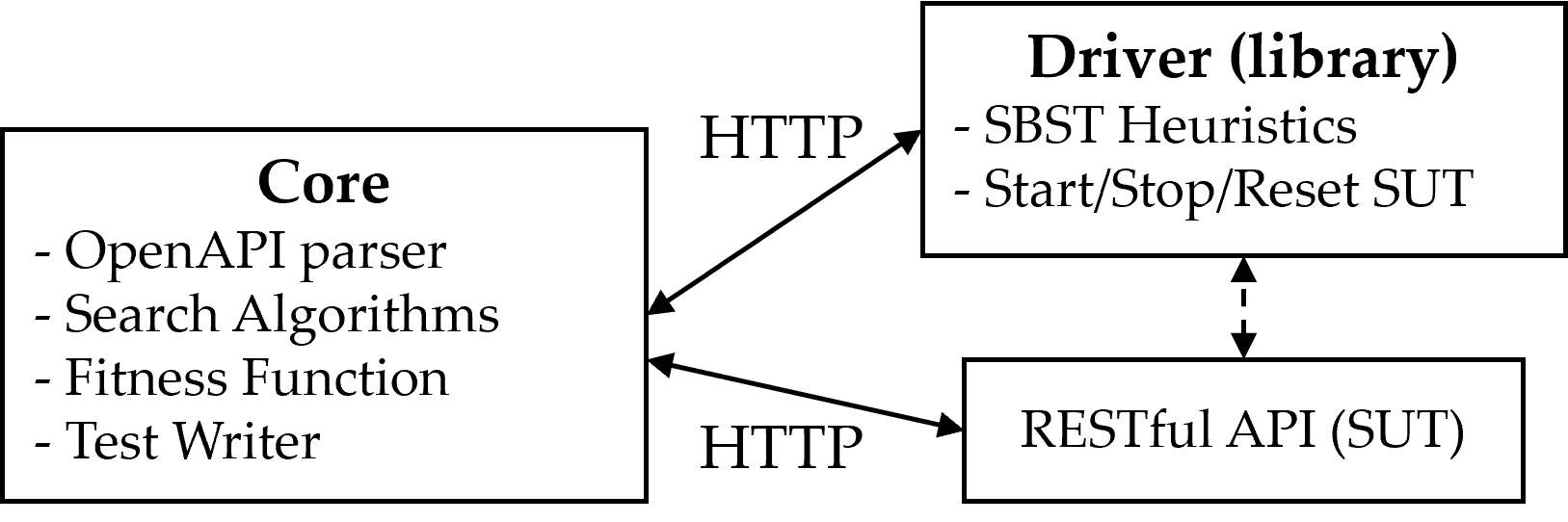}
	\caption{Architecture of \evo}
	\label{fig:evomaster}
\end{figure}

\evo is an open-source tool for fuzzing enterprise applications with search-based techniques, in the context of both white-box and black-box testing~\cite{arcuri2018evomaster,arcuri2020blackbox,arcuri2021evomaster}.
To enable white-box testing, the tool is composed of two components, i.e.,  \textit{driver} and \textit{core}, as shown in Figure~\ref{fig:evomaster}.
The \textit{driver} is responsible for collecting Search-Based Software Testing (SBST) heuristics with code instrumentation (currently targeting JVM~\cite{arcuri2018test} and JavaScript~\cite{js2022}) and controlling the SUT (i.e., start/stop/reset). 
It is implemented as a library, so that it is easy to be applied by the SUT using for example Maven and  Gradle. 
The \textit{core} encompasses the main functionality of generating test cases with search-based techniques, e.g., various search algorithms and fitness function.

To generate more effective white-box tests for enterprise APIs, \evo is equipped with a set of novel techniques.
For instance, \textit{boolean flag} is a common problem in handling white-box testing with search, i.e., no gradients for search to solve a constraint which is either \texttt{true} or \texttt{false}.
To enable fitness gradients for such problems in the source code, 
\evo is integrated with \textit{testability transformations}~\cite{arcuri2020testability,arcuri2021enhancing}. 
This enables \textit{branch distance} computations for such flag problems by transforming source code (e.g., via \textit{replacement} methods) with code instrumentation.
The \textit{replacement} methods also track inputs (referred as \textit{taint analysis}) for providing additional information to the search for solving this kind of problem.
In addition, enterprise APIs typically interact with databases.  
Database with various data would represent various states of the SUT.
To test the APIs with various states,  
\textit{SQL-handling}~\cite{arcuri2019sql,arcuri2020handling} was developed in \evo that can extract SQL queries and calculate heuristics for the queries at runtime.
Then, with these heuristics, \evo can directly generate data into the database (with SQL  commands such as \texttt{INSERT}).
REST is one of popular architectural styles for building web services.
To better support it, \evo employs a set of techniques designed in particular for the REST domain, e.g., OpenAPI parser, smart sampling, test reformulation for REST APIs~\cite{arcuri2019restful}, \textit{resource- and dependency- based strategies}~\cite{zhang2021resource,zhang2021enhancing}.

Furthermore,  \evo is enhanced with \textit{adaptive hypermutation}~\cite{zhang2021adaptive}.
It is as an advanced search mutator for handling long and structured chromosomes, like the tests for REST APIs which comprise a set of \texttt{INSERT} commands and a sequence of HTTP requests with query parameters and body payloads (e.g., JSON objects).

To serve as a SBST fuzzer, \evo includes the implementation of different search algorithms, i.e., MOSA~\cite{dynamosa2017}, WTS~\cite{GoA_TSE12, rojas2016detailed} and Random.
MIO~\cite{mio2017,arcuri2018test} is a search algorithm  
that was designed specifically for system test generation in the context of white-box testing.
MIO has been empirically studied by comparing with the other existing work (e.g., WTS, MOSA and Random), using artificial and real case studies.
Results showed that MIO achieved the overall best performance~\cite{mio2017,arcuri2018test}
when applied on the problem of fuzzing RESTful APIs.

\section{Related Work}
\label{sec:related}

To the best of our knowledge, there does not exist any technique for fuzzing modern RPC-based APIs (e.g., using frameworks like Apache Thrift, gRPC and SOFARPC). 
In addition, \evo seems the only open-source tool which supports white-box testing for Web APIs, and it gives the overall best results in recent empirical studies comparing existing fuzzers for REST APIs~\cite{kim2022arxiv,zhang2022open}.
However, currently \evo only supports fuzzing RESTful APIs~\cite{arcuri2019restful} and GraphQL APIs~\cite{belhadi2022graphql}.

In the literature, there has been work on the fuzzing of other kinds of web services.
The oldest approaches deal with black-box fuzzing of SOAP~\cite{curbera2002unraveling} web services, like for example~\cite{
tsai2002coyote,
offutt2004generating,
xu2005testing,
bai2005wsdl,
sneed2006wsdltest,
martin2006automated,
hanna2008fault,
ma2008wsdl,
bartolini2009ws,
li2016generating}.
SOAP is a type of RPC protocol. 
However, SOAP's reliance on XML format for schema definitions and message encoding has lead this protocol to lose most of its market share in industry (i.e., apart from maintaining legacy systems, it is not used so much any more for new projects). 

In recent years, there has been a large interest in the research community in testing RESTful APIs~\cite{golmohammadi2022testing,fielding2000architectural},
which are arguably the most common type of web services.
Several tools for fuzzing RESTful APIs have been developed in the research community, like for example (in alphabetic order) \bBOXRT~\cite{laranjeiro2021black}, \evo~\cite{arcuri2017restful}, \RESTest~\cite{restest2020}, \RestCT~\cite{wu2022icse}, \RESTler~\cite{restlerICSE2019} and \RestTestGen~\cite{viglianisi2020resttestgen}.
Another recently introduced type of web services is GraphQL~\cite{GraphQLFoundation}, which is getting momentum in industry.
However, there is only little work in academia on the automated testing of this kind of web services~\cite{vargas2018deviation,karlsson2020automatic,belhadi2022graphql,graphql2022arxiv}.

The automated testing of different kinds of web services (e.g., modern-RPC, SOAP, REST and GraphQL), share some common challenges (e.g., how to define white-box heuristics on the source code of the SUT, and how to deal with databases and interactions with external services).
However, there are as well specific research challenges for each type of web service, as we will show later in the paper.
A fuzzer for SOAP or REST APIs would not be directly applicable to a RPC web service, and vice-versa. 

In the literature, there are many applications of scientific research on the automation of software testing~\cite{bertolino2007software,ali2009systematic,godefroid2020fuzzing}.
Popular examples are AFL~\cite{AFL} for parsers and Sapienz for mobile applications~\cite{mao2016sapienz}.
Albeit possible, extending these kinds of tools from other testing domains for fuzzing RPC APIs would likely require major engineering and scientific effort.
Other domains like fuzzing network protocols (e.g., AFLNet~\cite{pham2020aflnet}) and
network devices (e.g., NDFuzz~\cite{zhang2022ndfuzz}) are closer to the fuzzing of Web APIs.
Still, non-trivial amount of work would be needed to adapt them to white-box fuzzing of RPC-based APIs.
For example, this could also explain why, to the best of our knowledge, none of these existing tools has been used so far to fuzz RESTful APIs, albeit their recent popularity in academia.

Given a client library for a RPC-based API, a unit test generator could be used directly on it, like for example the popular {\sc EvoSuite}~\cite{fraser2011evosuite} for Java classes.
This might work if the SUT and the client library are run in the same JVM.  
However, all the issues when dealing with system testing of web services would still be there, e.g., how to deal with databases and what to use as test oracle. 
Also, likely such unit testing tools would need some modifications (e.g., to collect coverage from all the classes and not just the RPC-client one). 
Therefore, how a unit test generator could be adapted and fare  in  such a system testing scenario is an open research question.

\section{Fuzzing RPC-based APIs}
\label{sec:fuzzing}

\begin{figure}
	\centering
	\includegraphics[width=0.9\linewidth]{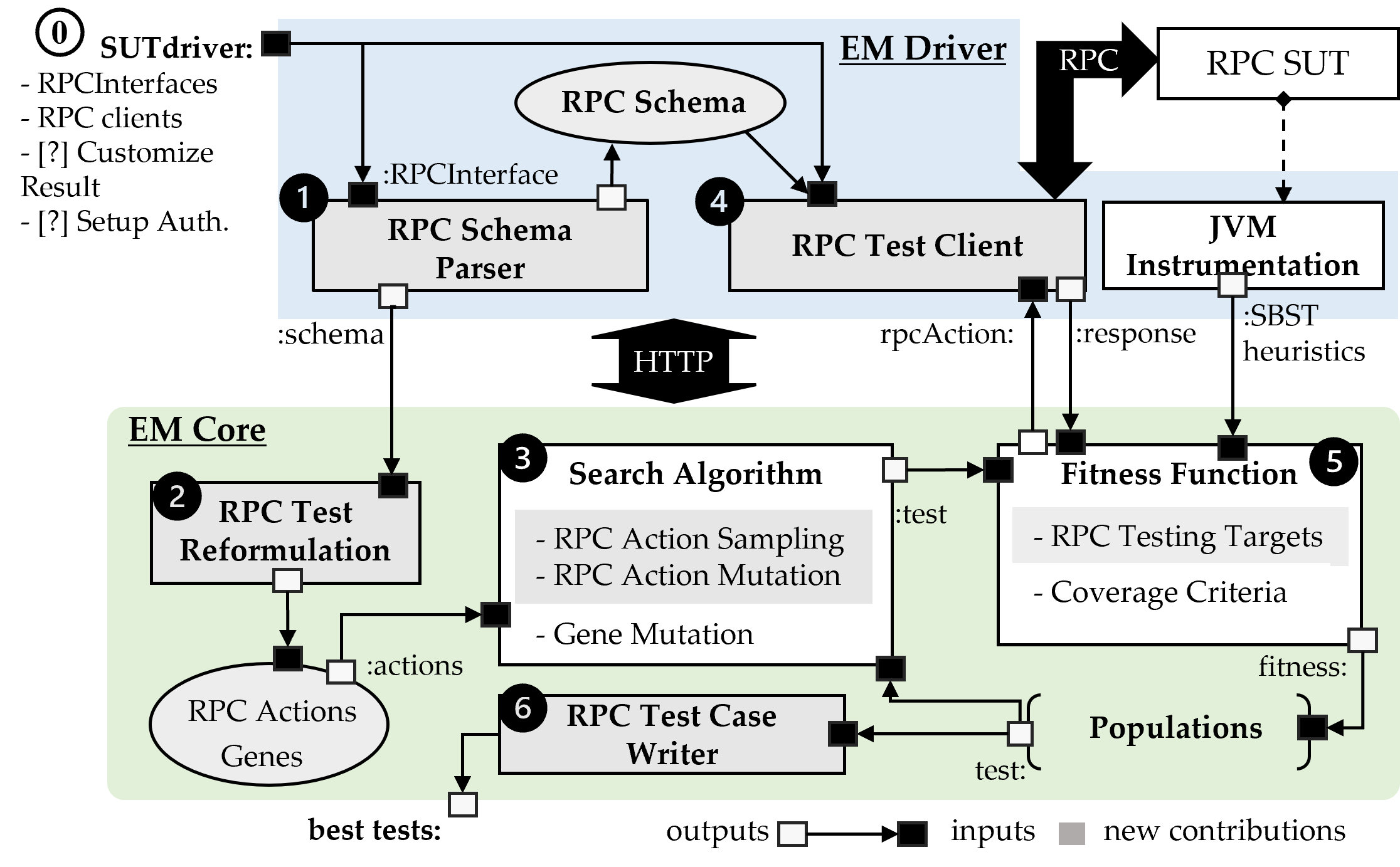}
	\caption{Overview of the approach built with \evo}
	\label{fig:approach}
\end{figure}

When addressing a new testing problem like the fuzzing of RPC-based APIs, several design decisions need to be made,
especially when using search-based techniques.
There is the need to specify the
\emph{search space} (Section~\ref{subsec:rpcschema}),
how to  represent the \emph{genotype} of an evolving individual (i.e., a test case in this context) (Section~\ref{sub:test-formulation}),
how to define the \emph{fitness function} to guide its evolution (Section~\ref{sub:fitness}),
which \emph{search operators} to employ to modify the evolving individuals (Section~\ref{sub:searcoperators}),
and how to \emph{output} the final results to the user (Section~\ref{sub:output}).

Building a fuzzer that can scale and be used on tens of industrial systems requires major engineering efforts,
throughout few years.
To evaluate the novel techniques presented in this paper, we did not start from scratch, but rather re-use and
extend an existing open-source fuzzer.
In particular, our novel approach is built on top of \evo (recall Section~\ref{sub:evomaster}).

Figure~\ref{fig:approach} represents an overview of our novel approach.
In order to fuzz RPC-based APIs, we purpose \textit{RPC Schema} specification which formulates necessary info to allow the execution of RPC function calls and the analysis of execution result.
In addition, with the specification,
as shown in the figure, the approach is composed of six steps distributed between the \textit{driver} and \textit{core} of \evo, plus initial settings manually provided by the user, for enabling automated fuzzing of RPC-based APIs with search techniques.
We briefly summarize these steps, where their details will be provided in the rest of this section.

To employ \evo, a \textit{SUTdriver} is required to be specified for implementing how to start/stop/reset the SUT (recall Section~\ref{sub:evomaster}).
In the context of RPC-based APIs testing, in the \textit{SUTdriver}, we further need the user to specify
(1) \textit{RPCInterfaces}: what interfaces are defining the API in the SUT with their class names
and
(2) \textit{RPC clients}: the corresponding client instances used to make RPC calls during test generation (Step~0).
Then, with the specified interface info, \textit{RPC Schema Parser} will extract and identify the API schema based on proposed \textit{RPC schema} specification, in order to access the RPC functions (Step~1). 
At the \textit{core} side, the extracted schemas will be further reformulated (Step~2) to be as  components (i.e., \textit{RPC Actions} and \textit{Genes}) of the search for producing tests (Step~3).
In our approach, a generated test is evaluated by its execution on the SUT (Step~4) performed on the \textit{driver} side.
Then, responses, SBST heuristics (e.g., code coverage with code instrumentation) and  identified potential faults resulted in the execution will be returned to the \textit{Fitness Function} (Step~5) for calculating the fitness value of the executed test. 
Producing and evaluating tests are performed iteratively (i.e., Steps~3-5), within a given search budget.
At the end of the search, a set of the best (in terms of code coverage and fault detection) tests for the RPC-based SUT will be outputted (Step~6) with a given format (e.g., JUnit 5).

\subsection{Search Space}
\label{subsec:rpcschema}

At a high level, a RPC-based API can be seen as process that opens a TPC/UDP port on a given host, and then reply to incoming
messages formatted with a given application-layer protocol.
Such protocol could vary among the different RPC implementations.
Furthermore, the API would reply only to requests for its defined methods, requiring the right number and type of input parameters.
This means that sending random bytes over the TCP/UDP connections would unlikely result in any meaningful response from the API,
and possibly no execution of the code of its business logic.

To address this issue, it would be important for a fuzzer to send well-formatted messages for the different remote APIs exposed by the web service.
Given a \emph{schema} that specifies which methods can be called, a fuzzer can then generate calls with the right input parameters.
Considering that these methods can take inputs complex data such strings, objects and arrays, the search space of possible
inputs is huge, even when using a schema to constrain what will be sent.
Only with some specific inputs faults could be revealed and code coverage optimized.
Furthermore, to test a specific endpoint, there might be the need to call a previous one to set the state (e.g., a database) of the
API.
This means that a test case would hence be a sequence of one or more remote calls towards the API, which increases the search space
even further.
To complicate this even further, to achieve higher code coverage the API might require setting up the \emph{environment} in which
it operates.
For example, advance fuzzers can also add data directly into SQL databases as part of an initialization phase, based on what queries the API executes on the database.
This further extends the search space of possible test cases that the fuzzers need to explore.

Nowadays, there exist various RPC frameworks for building modern RPC-based APIs, e.g., Thrift~\cite{thrift}, gRPC~\cite{gRPC}, Dubbo~\cite{dubbo} and
SOFARPC~\cite{sofarpc}.
As discussed in Section~\ref{sub:rpcbg}, most of the techniques would result in 
\textit{RPCInterface}s (e.g., implemented as  \texttt{interface} or \texttt{abstract class})
in their API implementations representing how the services can be accessed, together with a client-stub to make the actual RPC calls.
Considering all the possible types of communication protocols supported by the different RPC frameworks, calling a RPC API directly from a fuzzer would be a major technical endeavor. 
Furthermore, it would require to support the different schema languages for each framework, like for example \texttt{.thrift} (see Figure~\ref{sub:thriftschema}) and \texttt{.proto} (see Figure~\ref{sub:gRPCschema}) formats,
and there would be limitations when
the schema file might not be available, such as the APIs implemented with SOFARPC and Dubbo.

In order to enable automating testing of RPC-based APIs in a more generic way, 
in this paper we propose a schema specification specific to RPC domain that formulates main concepts for facilitating invocations of RPC function calls  and result analysis.
Such a specification can be automatically extracted based on \textit{RPCInterface}s,
regardless of which RPC framework is employed by the API.
This schema defines the \emph{search space} for the fuzzing, as we will evolve test cases complying with such schema.
Then, we employ the actual client-libraries of the APIs to make the RPC calls.

\subsubsection{RPC Schema Specification}
\label{subsubsec:rpcinterface}

\begin{figure}
	\centering
	\includegraphics[width=1\linewidth]{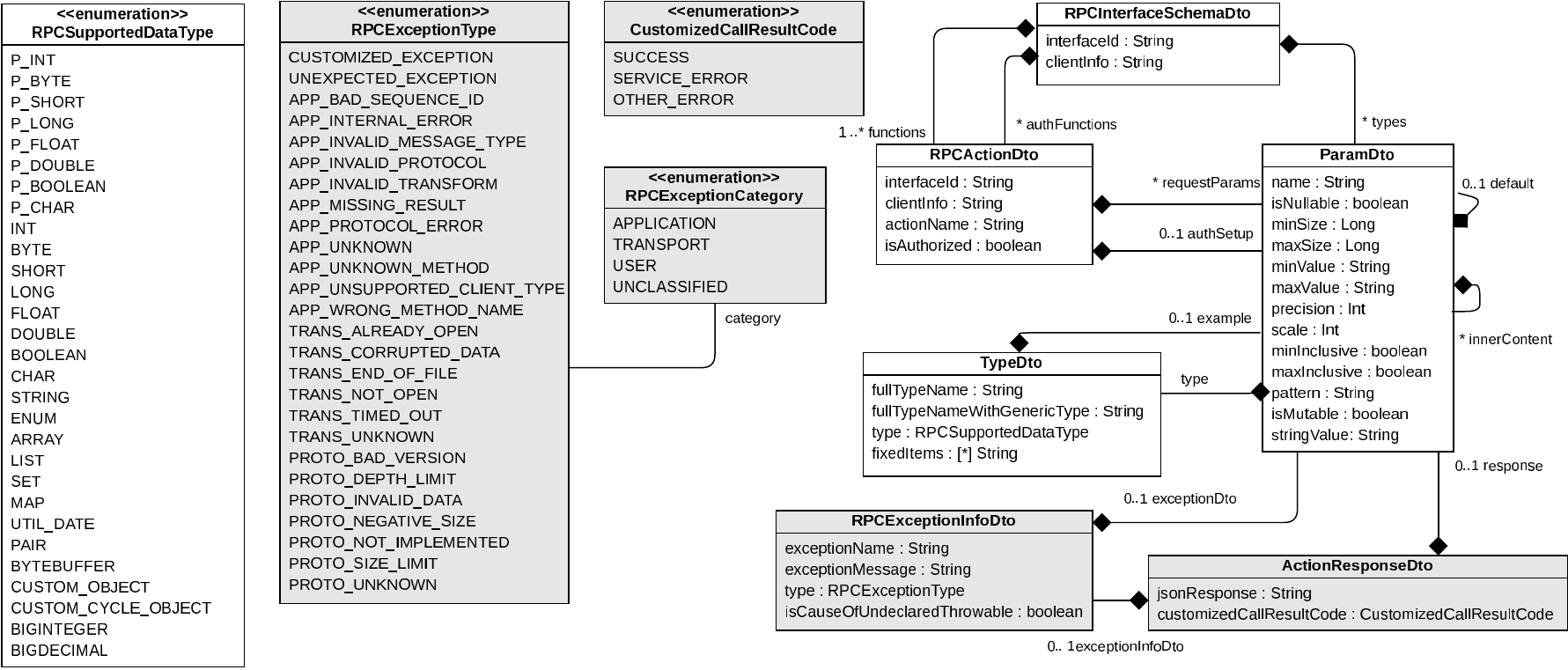}
	\caption{Data Transfer Objects defined in our RPC schema specification}
	\label{fig:rpcdto}
\end{figure}

Our RPC Schema is defined with a Data Transfer Object (DTO), which can then be instantiated in different formats,
like for example JSON.
Figure~\ref{fig:rpcdto} shows our RPC schema specification with an UML class diagram.

To extract info for enabling invocations of RPC function calls,
there exist five main concepts to define \textit{RPCInterfaces} (denoted as classes with white background in Figure~\ref{fig:rpcdto}):
\begin{itemize}
	\item \rpcschemadto: it represents the \textit{RPCInterface}, such as the \texttt{Interface} with Thrift (see Figure~\ref{sub:thriftjava}) and \texttt{abstract class} with gRPC (see Figure~\ref{sub:gRPCjava}). 
	A \rpcschemadto comprises one or more \rpcfunctiondto (see \textit{1..* functions}), a set of functions for authentication handling (see \textit{* authFunctions}) and a set of specifications of data types (see \textit{* types}).
	For instance, \texttt{NcsService.Iface} interface has a \texttt{bessj} function and employs \texttt{Dto} data structure (as shown in Figure~\ref{sub:thriftjava}). 
	Note that a RPC-based API might exist multiple interfaces as industrial APIs which we studied in this paper.
	
	\item \rpcfunctiondto: it captures info to make a RPC function call, i.e., input parameters if exist (see \textit{* requestParams}) and additional authentication setup (see \textit{0..1 authSetup}).
	Each \rpcfunctiondto also has \textit{interfaceId}, \textit{clientInfo} and \textit{actionName} properties to identify the RPC function to call.
	In addition, we identify a property \textit{isAuthorized} representing whether the \rpcfunctiondto is restricted with authentications in its implementation.
	
	\item \rpcparamdto: it is used to describe values of input parameters and return.
	A \rpcparamdto links to an explicit datatype (see \textit{type}) and might be composed of a set of  {\rpcparamdto}s for representing complex data types, such as object, collection and map (see \textit{* innerContent}).
	The \rpcparamdto might be specified with a default value (see \textit{0..1 default}), e.g., a field in a DTO can be assigned with a default value. 
	In addition, we define \textit{stringValue} to assign a value for the input parameter or represent the actual value of the return.
	Note that \textit{stringValue} is applicable only if there is no any internal elements.
	To construct constraints of the input parameters if exist, we define a set of properties in \rpcparamdto as:
		\begin{itemize}
			\item \textit{isNullable} represents whether the parameter is nullable to make the call.  
			
			\item \textit{isMutable} indicates whether the parameter is mutable. 
			A value of the property is derived based on whether the parameter is assigned with a fixed value.
			For instance, a parameter must be \texttt{true} if it is specified with \texttt{@AssertTrue}, thus the parameter is considered as immutable.
			 
			\item \textit{minSize} and \textit{maxSize} represents boundaries in size if specified.
			The constraint could be applicable to data types, i.e., collection, map, array and char sequence (e.g., string).
			
			\item \textit{minValue} and \textit{minInclusive} are used to represent a minimum value, and a value of the parameter must be higher than or equal to the minimum. 
			
			\item \textit{maxValue} and \textit{maxInclusive} are used to represent a maximum value, and a value of the parameter must be lower than or equal to the maximum. 
			
			\item \textit{precision} and \textit{scale} capture constraints for numeric values regarding its precision and scale (e.g., number of digits in their decimal part).
			
			\item \textit{pattern} represents a regular expression that a string value must match.
			
		\end{itemize}
	Such captured constraints could contribute to test data generation for fuzzing Web APIs, by sampling values within the boundaries of these constraints.  
	Values of all of the constraint properties could be derived automatically based on the \textit{PRCInterface}, which is explained in the \textit{RPC schema extraction} (see Section~\ref{subsec:rpcextraction}).
	
	\item \rpctypedto and \rpcdatetype identify the datatype info of the \rpcparamdto. 
	A list of data types we support is defined as an enumeration \rpcdatetype which covers the most commonly used data types, i.e., 
	\textit{array}, 
	\textit{byte buffer}, 
	\textit{date}, 
	\textit{enumeration}, 
	\textit{list}, 
	\textit{map}, 
	\textit{set}, 
	\textit{string}, 
	\textit{integer}, 
	\textit{boolean},
	\textit{double},
	\textit{float},
	\textit{long}, 
	\textit{character}, 
	\textit{byte}, 
	\textit{short}, 
	\textit{big integer},
	\textit{big decimal},
	and any customized DTO object, for enabling the fuzzing of RPC-based APIs.
	In \rpctypedto, it can be specified with an example (see \textit{0..1 example}) for representing a generic type of collection, array and map. 
	Note that this list of supported data types is not meant to be complete for all RPC frameworks.
	But, if needed, it can be extended.
	
\end{itemize}

\subsubsection{RPC Schema Extraction and Execution Support}
\label{subsec:rpcextraction}

As a white-box fuzzer, beside source code of SUT,
a \textit{SUTdriver} is the only input which \evo needs a user to specify (recall Section~\ref{sub:evomaster}).
Then, the \textit{SUTdriver} is employed at the \textit{driver} side for, e.g., starting/stopping/resetting the SUT.
In the context of RPC-based API fuzzing, we further need the user to provide info of \textit{RPCInterface}s and corresponding client instances for extracting the API schema and accessing the SUT.
As shown in Figure~\ref{fig:approach}, in the \emph{driver}, with the provided \textit{SUTdriver} (Step~0),
we developed a \textit{RPC Schema Parser}, by directly extracting the interface definitions (which do represent the API schema) from the source code using \textit{reflection technique},
such as Java Reflection\footnote{https://www.oracle.com/technical-resources/articles/java/javareflection.html}.
Thus, with any RPC framework, if the available RPC functions are defined as an \texttt{interface}/\texttt{abstract class} (which is usually the case), our approach could be applicable.
The extracted information is further formulated as a generic \textit{RPC Schema} (see Section~\ref{subsec:rpcschema}), i.e., a \textit{RPCInterface} will be formulated as a \rpcschemadto which contains specifications to invoke RPC function calls (i.e., \rpcfunctiondto (Step 1 $\rightarrow$ Step 2)).
In addition, we developed \textit{RPC Test Client} which allows to make a RPC function call against the SUT with \rpcfunctiondto, then return \rpcresponse (Step 5 $\leftrightarrow$ Step 4) using specified RPC client instances.
The \textit{driver} is implemented as a service using REST API, and the two components (i.e., \textit{RPC Schema Parser} and \textit{RPC Test Client}) are exposed as two HTTP endpoints, i.e., \texttt{/infoSUT} for extracting RPC API schema and \texttt{/newAction} for executing RPC function calls.
Thus, with a provided \textit{SUTdriver}, our \textit{driver} employed with proposed \textit{RPC Schema} would allow a unique interface of our tool to support invocations of RPC functions and result analysis.
This is an essential prerequisite for fuzzing RPC-based API.

Note that instead of enabling RPC function execution at the \textit{driver} side, an alternative approach would have been to include the two components and RPC API client-library directly into \emph{core} process, which might be more efficient (as calls from the \emph{core} do not need to go through the \emph{driver} with HTTP requests).
But that would introduce a lot of \emph{usability} issues to configure it up (e.g., how to dynamically load a library at runtime, and how to deal with different JVM versions and different programming languages).
When introducing a novel approach, it is important to take into account how complex it is to set it up by practitioners.
For this, industry collaborations, where actual engineers use these techniques on their systems (as we do for this paper), are paramount.

\subsubsection{SUTdriver Implementation}

\begin{figure}
\begin{lstlisting}[language=java,basicstyle=\scriptsize,escapechar=©]
public class EmbeddedEvoMasterController extends EmbeddedSutController {

	private ConfigurableApplicationContext ctx;
	private NcsService.Client client;
	private TTransport transport;
	private TProtocol protocol;

	@Override
	public String startSut() {©\label{line:startSutStart}©

		ctx = SpringApplication.run(NcsApplication.class, new String[]{
			"--server.port=0"
		});

		String url = "http://localhost:"+getSutPort()+"/ncs";

		try {
			transport = new THttpClient(url);©\label{line:clientStart}©
			protocol = new TBinaryProtocol(transport);
			client = new NcsService.Client(protocol);©\label{line:clientEnd}©
		} catch (TTransportException e) {}

		return url;
	}©\label{line:startSutEnd}©

	@Override
	public ProblemInfo getProblemInfo() {

		return new RPCProblem(new HashMap<String, Object>() {{©\label{line:rpcinterfaceStart}©
				put(NcsService.Iface.class.getName(), client);
		}});©\label{line:rpcinterfaceEnd}©
	}

	@Override
	public CustomizedCallResultCode categorizeBasedOnResponse(Object response) {return null;}©\label{line:customizeResult}©

	@Override
	public List<AuthenticationDto> getInfoForAuthentication() {return null;}©\label{line:authInfo}©

	@Override
	public List<CustomizedRequestValueDto> getCustomizedValueInRequests() {return null;}©\label{line:authInRequest}©

}
\end{lstlisting}
\caption{Snippet code of a driver for NcsService (see Figure~\ref{fig:thriftexample})}
\label{fig:driver}
\end{figure}

Figure~\ref{fig:driver} represents an example of a \textit{SUTdriver} for manipulating the SUT and specifying info of a RPC-based API.
For instance, a \texttt{startSut} method at lines~\ref{line:startSutStart}--\ref{line:startSutEnd} represent how to start a RPC-based SUT which is implemented with the Thrift framework and SpringBoot.
The method also instantiates needed clients to access the SUT after it starts, see lines~\ref{line:clientStart}--\ref{line:clientEnd}.
To provide info specific to RPC problem, lines~\ref{line:rpcinterfaceStart}--\ref{line:rpcinterfaceEnd} specify the \textit{RPCInterface} (i.e., \texttt{NcsService.Iface} see Figure~\ref{sub:thriftjava}) and corresponding client instance.
Note that the info is specified with a map since an API might have multiple \textit{RPCInterface}s as we observed in our industrial case studies.

In addition,
each framework or each company might define their own rules to represent results.
For instance, we found that,
in our industrial case study, in most cases a failed function call would not result in any exception thrown to avoid propagation of exceptions in the distributed system, since the services are connected with each other.
Thus, inside the response, our industrial partner has its own customized specification to reflect results of RPC function calls that are linked to their business logic.
Without a thrown exception, a response representing an error might be falsely identified as a success if no further info is provided.
To address this concrete issue in industrial APIs, in our approach, we provide an extensible method (i.e., \texttt{getCustomizedValueInRequests}  at line~\ref{line:customizeResult}) to enable customized categorization of responses with the three levels as \rpccustomizedresult defined in our \textit{RPC schema} (Section~\ref{subsec:rpcschema}).
By extending the method, the user could directly link their own rules into our testing context.
Note that such setup can be easily reused by multiple SUTs if they use same customized specification (as it was for all web services developed by our industrial partner).

As an enterprise system, authentication is typically required to be handled.
However, there are many different ways to implement an authentication system in a RPC API, as it is usually not supported natively (at least not in Thrift).
For this paper, we are mainly supporting the authentication systems used by our industrial partner.
Authentication tokens need to be sent as a field in payloads of the messages (similarly as HTTP authentication headers in RESTful APIs).
An authentication token can be either \emph{static} (i.e., pre-fixed) or \emph{dynamic}.
This latter requires to get the token from an endpoint (e.g., a \texttt{login} RPC endpoint where valid username/password info must be provided), and then add it to all following RPC calls.
In our implementation, we support both approaches, which needs to be configured in the \emph{driver}, i.e., by extending the method \texttt{getInfoForAuthentication} at line~\ref{line:authInfo} and \texttt{getCustomizedValueInRequests} at line~\ref{line:authInRequest} as shown in Figure~\ref{fig:driver}.
To serve a more fine-tuned setup for authentication, we enable options to specify (1) if either the authentication is applied for all API functions;  or (2) specific only to some functions in that SUT, that could be filtered by names  or by special annotations applied on these functions.
More detail about how to configure the option could be found in two DTOs, i.e., \texttt{JsonAuthRPCEndpointDto} and \texttt{CustomizedRequestValueDto}, in our implementation\textsuperscript{\ref{foot:link}}.

\subsubsection{RPC Schema Parser}

Regarding the extraction of RPC interface definitions, currently, we target JVM RPC-based APIs using Java Reflection.
As examples shown in Figures~\ref{fig:thriftexample} and \ref{fig:gRPCexample}, a client-stub \textit{RPCInterface} is composed of a set of available RPC functions to be extracted.
Each operation in the interface depicts a RPC function to be called in this service.
Then, with reflection, for each interface, we identify all such public methods, and then further extract info on their input parameters, return type and declared exception types.

Regarding datatype, as currently targeting JVM projects, we have supported  the most commonly used data types, i.e.,
\texttt{Array},
\texttt{ByteBuffer},
\texttt{Date},
\texttt{Enum},
\texttt{List},
\texttt{Map},
\texttt{Set},
\texttt{String},
\texttt{Integer}, \texttt{int},
\texttt{Boolean}, \texttt{bool},
\texttt{Double}, \texttt{double},
\texttt{Float}, \texttt{float},
\texttt{Long}, \texttt{long}
\texttt{Character}, \texttt{char}
\texttt{Byte}, \texttt{byte},
\texttt{Short}, \texttt{short},
\texttt{BigInteger},
\texttt{BigDecimal},
and any customized DTO object.
For the handling of generics,
we support their instantiations for any of these common data types.
Note that
all of the datatype could be mapped to an item defined in \rpcdatetype.

Regarding the parameter, besides its datatype, we also need to extract info, such as accessibility and constraints if they exist.
Extracting accessibility is needed for the parameter typed with DTO, then its fields might be publicly accessible or not, i.e., declared as \texttt{public} or not in Java.
If the filed is not publicly accessible, there is a need to further extract its existing getter and setter that would be used in assertion generation (with getter) and parameter construction (with setter) in our context.
Note that
the accessibility info for each parameter is maintained inside the \textit{RPC Test Client} that does not expose in DTO, since the user does not need to care about how to construct the data instance for the parameter and assertion generation.
More detail about how the info is constructed can be found in the class \texttt{AccessibleSchema}\textsuperscript{\ref{foot:link}}.

Regarding the constraints, a parameter might be specified with constraints in its implementation.
For example, an integer representing the day of the month could be constrained between the values 1 and 31.
To make RPC function calls that do not fail due to input validation, we need to handle such constraints when generating input data for the call.
Therefore, for each parameter, with the proposed schema, we define possible constraints as properties of \rpcparamdto (see Section~\ref{subsubsec:rpcinterface} and Figure~\ref{fig:rpcdto}).
With the extraction, we further identify the properties based on the data types.
For instance, all parameters are defined with a property named \textit{isNullable} representing whether a parameter object
can be \textit{null} (the value of this property for all primitive types is always false).
Parameters with numeric data types are defined with \textit{min} and \textit{max} properties.
For parameters representing collections (e.g., maps and lists) and string types, properties for constraining their size/length are defined, i.e., \textit{minSize} and \textit{maxSize}.
For strings, we define \textit{pattern} for supporting a constraint specified with regular expressions.
If a string has to represent a numeric value,
we use \textit{minValue} and \textit{maxValue} for supporting a possible range constraint for it.

To identify constraints defined in the interface definitions (typically with annotations), we enable constraint extraction on \texttt{javax.\-validation.\-contraints}~\cite{javaxvalidationconstraints} which is the standard library for defining built-in constraints for Java objects.
We support 16 commonly used constraints, i.e.,
\texttt{AssertFalse},
\texttt{AssertTrue},
\texttt{DecimalMax},
\texttt{DecimalMin},
\texttt{Digits},
\texttt{Max},
\texttt{Min},
\texttt{Negative},
\texttt{NegativeOrZero},
\texttt{NotBlank},
\texttt{NotEmpty},
\texttt{NotNull},
\texttt{Pattern},
\texttt{Positive},
\texttt{PositiveOrZero},
and \texttt{Size}.
Besides standard \texttt{javax} annotations, constraints could be defined in other ways as well.
For instance, in Thrift, whether a field is \textit{required} is represented by a \texttt{requirementType} property of the \texttt{FieldMetaData} class.
Thus, in order to deal with constraints in the Thrift framework, we further extract and analyze the  \texttt{metaDataMap} object in the interface for obtaining such constraints.

In addition, since there is no general standard to restrict such interface implementations (as long as it compiles),
the method and the data type might use Java Generics (as we found in our industrial case study).
Therefore, we further handle such generic types when processing RPC function extraction, e.g., analyze \texttt{getParameterizedType} for each parameter.

With the \textit{RPC Schema Parser}, it is capable of formulating each \textit{RPCInterface} as \rpcschemadto shown in Figure~\ref{fig:rpcdto}.

\subsection{Genotype Representation}
\label{sub:test-formulation}

Given an extracted \textit{RPCSchemaDto} schema, we need to define how to represent the genotype of the evolving test cases.
In our context, a test could be reformulated as an individual which is composed of a sequence of RPC function calls.
Each function call is formulated as \textit{RPCCallAction}, which comprises
the method name,
input parameters (if any), optional authentication info, and a response (if declared).

For each input parameter, we define a \emph{gene} with a specific type to represent the parameter.
A gene is an instance for the specific type, with constraints on how it can be \emph{mutated} (i.e., modified by the search
operators) during the search.
For example, a numerical parameter could be internally represented as an integer, initialized with a random value, where
the search operators could add or move a delta from such value during its evolution.
Textual parameters could be represented with a string, where search operators can either modify its characters, and add or delete
some of them (and so changing the length of the string).
For more complex types, genes can be hierarchically combined in a tree structure.
For example, an object is represented with a gene that has one child gene for each field of the object (and so on recursively,
if any of these child fields is an object itself).

There are many types of possible parameters to handle.
To achieve a full support to be able to handle RPC-based APIs, we re-use (and extended where needed) the gene system already
present in our \evo fuzzer.
Regarding the input parameters, we could re-use existing \textit{Gene} objects already defined in \evo for supporting REST API testing, like for example:
\begin{itemize}
	\item 	
	 Straightforward mapping: \textit{ArrayGene} for \texttt{Array}, \texttt{Set} and \textit{List};
	\textit{BooleanGene} for \texttt{Boolean} and \texttt{bool};
	\textit{DoubleGene} for \texttt{Double} and \texttt{double};
	\textit{LongGene} for \texttt{Long} and \texttt{long};
	\textit{FloatGene} for \texttt{Float} and \texttt{float};
	\textit{EnumGene} for \texttt{Enum};
	\textit{DateGene}, \textit{DateTimeGene} and \textit{TimeGene} for \texttt{DateTime};
	\item	  
	\textit{MapGene} for \texttt{Map}. Note that the original version of \textit{MapGene} only supports key with string type.
	However, other types such as enum and integer are quite common in RPC-based APIs.
	Therefore, we further extended \textit{MapGene} for enabling key to be specified with \textit{IntegerGene}, \textit{StringGene}, \textit{LongGene} and \textit{EnumGene}.
	\item 
	\textit{IntegerGene} for \texttt{Integer}, \texttt{int}, \texttt{Short}, \texttt{short}, \texttt{Byte} and \texttt{byte} (various types here are distinguished by min value and max value, e.g., max value is configured as 127 for \texttt{Byte} by default if it is not constrained);
	\item 
	
	\textit{StringGene} for \texttt{Character}, \texttt{char}, \texttt{String} and \texttt{ByteBuffer}  (various types are distinguished by min and max length, e.g., max length is configured as 1 for \texttt{char} by default if it is not constrained);
	\item 
	\textit{RegexGene} for a pattern specified in \texttt{String} parameter;
	\item
	\textit{ObjectGene} for representing customized class object;
	\item 
	\textit{CycleObjectGene} for a field in the customized class object that leads to a cycle;
	\item 
	\textit{Optional} is for handling any parameter whose \textit{isNullable} property is true.
\end{itemize}

In addition, we also purpose new genes, such as \textit{BigDecimalGene} and \textit{BigIntegerGene} for \texttt{BigDecimal} and \texttt{BigInteger} respectively. 
In the original implementation of \textit{Gene}s in \evo, constraints for all types are not fully supported.
Therefore, to fully support  testing the RPC APIs in our case study, we extended genes by enabling all constraints we defined in \textit{RPC schema}, such as handling precision and scale for numeric genes, and min and max size constraints for \textit{ArrayGene}, \textit{MapGene} and \textit{StringGene}.
This means that, when these genes are either sampled at random, or modified throughout the search via mutation operators,
all (linear) constraints are kept satisfied (e.g., a mutation operator would not try to increase a numeric value if it is already
at its maximum as defined in its gene constraints).

To test a RPC-based API, the input parameters could be either automatically generated or manually configured by the user 
(e.g., unlike header in HTTP request, in RPC function call, authentication info could be specified as parts of input DTOs).
The former one would be handled by search techniques in our approach.
The way to enable authentication  as part of the input parameters can be identified as the latter option, i.e., manually defined inputs.
To allow further combinational handling with both automatic and manual solutions, we decided to extend the test reformulation with a new gene, i.e., \textit{SeededGene}, for handling manual inputs in a more generic way.
A \textit{SeededGene}, representing a gene which has a set of candidates, is constructed with: 
(1) \textit{gene} is the original genotype of the parameter that could be mutated with the search;  
(2) \textit{seeded} is an \textit{EnumGene} with the same type as \textit{gene} presenting enumerated candidates; 
and (3) \textit{employSeeded} is a boolean to indicate whether the original \emph{gene} or the \textit{seeded} gene is used for the phenotype of \textit{SeededGene}.
Besides handling authentication info, this kind of the gene also allows further seeding with existing data (if any) that would be in particular useful in solving industrial problems. 

To be able to efficiently fuzz real-world APIs, currently \evo has more than 80 different types of genes in its search-based fuzzer engine~\cite{zenodo150evomaster}.
A full description of each of them is not viable here.
For low-level technical details, the interested reader can check out our implementation~\cite{zenodo150evomaster},
in particular the code under the \texttt{org.evomaster.core.search} package.

\subsection{Fitness Function}
\label{sub:fitness}

To evaluate the fitness of a test case, we need to be able to make calls toward the API, with the right inputs, in the right format.
The fitness itself will be based on two different kinds of metrics:
white-box heuristics based on the execution in the source code (which requires the API to be instrumented with probes),
and black-box heuristics based on the responses returned from each RPC call.

To make calls on the API, we use the \emph{client library} provided by the API itself (recall Section~\ref{subsec:rpcextraction}).
Most RPC frameworks (e.g., Thrift and gRPC) provide ways to automatically generate client libraries (recall Section~\ref{sub:rpcbg}).
However, there would be several technical issues in dynamically loading such library inside the core process of \evo.
Our solution is to let the user to specify (and link) such client libraries in the \emph{driver} classes that need to
be written to run white-box mode of \evo (recall Section~\ref{sub:evomaster}).
This means that, when a test case needs to be evaluated, the \emph{core} process sends a representation (in JSON format) of such a test case to the \emph{driver}, and then the driver executes the actual RPC call and collects its response.
Plus, the driver also collects any white-box heuristics from the instrumented API.
Then, all this information is sent back to \emph{core}, where the fitness value for the test is computed.

This architecture to support fuzzing of RPC APIs introduces some latency, as the \evo core does not communicate directly with the API.
However, it has major benefits, as it is much easier to setup and implement (e.g., there is no need to parse
any \texttt{.thrift} or \texttt{.proto} file), as well as enabling supporting all different kinds of RPC frameworks with little effort.

\subsubsection{RPC Execution Result Analysis}
\label{subsubsec:resultanalysi}

By using client to invoke RPC function call, a result received at the client side could be a return value as defined or an exception thrown from the API.
To enable the result analysis, we proposed five main concepts denoted as classes with gray background in Figure~\ref{fig:rpcdto}, i.e., \rpcresponse, \rpcexceptioninfodto, \rpcexceptioncategory, \rpcexceptiontype and \rpccustomizedresult.
\rpcresponse is a DTO which captures all info returned from a RPC function call, i.e., throw an exception (see \textit{0..1 exceptionInfoDto}) or return a value as specified (see \textit{0..1 response}).

Regarding exception,
handling the exception info for  RPC functions is crucial for testing purposes, e.g., to be able to use automated oracles to identify faults in the SUT.
To analyze an exception, in our proposed schema,
we define \rpcexceptioninfodto which captures \textit{exceptionName}, \textit{exceptionMessage}, \textit{type} and \textit{exceptionDto}, which is an optional DTO representing possible additional info for customized exceptions (e.g., the exceptions declared with the keyword \texttt{throws} in Java).
In addition, when invoking RPC function calls with clients which could be proxy clients,  an exception caught at the client side might be wrapped, such as \texttt{UndeclaredThrowableException}\footnote{\url{https://docs.oracle.com/javase/8/docs/api/java/lang/reflect/InvocationHandler.html}} in Java.
To get the exact exception info,
we further extract and analyze the actual exception (e.g., with \texttt{cause} of \texttt{UndeclaredThrowableException}) as \rpcexceptioninfodto.
We also perform further exception analysis on \texttt{UndeclaredThrowableException}, as it was needed for our industrial case study, and the property \textit{isCasueOfUndeclaredThrowableException} represents whether such a wrapped exception is thrown from the SUT.
Note that the actual exception analysis could be extended in future when needed.

Beside \textit{exceptionName} and \textit{exceptionMessage},
to better identify exceptions in the context of RPC-based APIs, based on domain knowledge, we classify exceptions into four categories as \rpcexceptioncategory: \textit{APPLICATION} (e.g., internal server errors), \textit{TRANSPORT} (e.g., connection timeouts), \textit{USER} (e.g., sending invalid data), and
\textit{UNCLASSIFIED}.
Different RPC frameworks can define their own exceptions for handling various situations for RPC (e.g., type of \texttt{TApplicationException}~\cite{TApplicationException}
defined in \texttt{TException} for Thrift, status~\cite{gRPCstatuscode}
defined in \texttt{StatusException} and \texttt{StatusRuntimeException} for gRPC).
To cover such knowledge captured in various RPC frameworks,
we define \rpcexceptiontype, and
each of the type should belong to a category in \rpcexceptioncategory.
The \rpcexceptiontype now provides a full support for analyzing exceptions in
the Thrift framework, which covers the complete 24 exception types from
\texttt{TApplicationException} (refer to \textit{APPLICATION} category),
\texttt{TProtocolException} (refer to \textit{USER} category), and
\texttt{TTransportException} (refer to \textit{TRANSPORT} category).
In addition,
we define two generic exception types, i.e., \textit{CUSTOMIZED\_EXCEPTION} representing a declared exception (e.g., \texttt{throws} keyword in Java), \textit{UNEXPECTED\_EXCEPTION} representing an exception which is not declared in the function and does not belong to any other identified types (e.g., \texttt{RuntimeException} in Java).
The generic exception types link to \textit{UNCLASSIFIED} category that covers the cases whereby the exception type is unspecified or its identification is not supported yet for linking it to a specific RPC exception (like Thrift).

With \rpcresponse,
considering how a RPC call is handled by the SUT and if there is any exception,
we classified it into seven kinds of execution results which would contribute to define search heuristics for optimizing generated tests:
\begin{itemize}
	\item
	(ER1)
	\textit{internal error}: an exception which represents an internal error is thrown, e.g., \texttt{TApplicatinException} with \texttt{INTERNAL\_ERROR} type in Thrift.
	\item
	(ER2)
	\textit{user error}: if an exception was thrown that can be traced to a failed input validation, based on Thrift's protocol errors. 
	\item
	(ER3)
	\textit{transport error}: an exception which represents transport errors is thrown.
	\item
	(ER4)
	\textit{other exception}: other exception (e.g., other types of \texttt{TApplication} except internal error) is thrown.
	\item
	(ER5)
	\textit{declared exception}: an exception declared in the function is thrown.
	\item
	(ER6)
	\textit{unexpected exception}: any other exception which is not declared in the function is thrown.
\item
	(ER7)
	\textit{handled}: a value is returned as declared without any exception thrown. If users specify their result categorization, this label  is further refined as one of \textit{success}, \textit{service error} and \textit{other error}.
\end{itemize}

With HTTP, a result for a request could be identified based on \textit{status code} in its response,
e.g., 2xx indicates a success, 4xx indicates a client error and 5xx indicates a server error.
Such a standard is useful in developing automated testing approaches, e.g., reward requests with 500 status code (for finding  potential faults in the SUT) and 2xx status code (for covering a successful request).
However, in the context of RPC, there does not exist such standard, and a result (e.g., success or failure) of the call cannot be directly determinate based on the return value if there is no  exception thrown.
Therefore, we propose \rpccustomizedresult which defines three categories (i.e., \textit{SUCCESS}, \textit{SERVICE\_ERROR} and \textit{OTHER\_ERROR}) to better identify a return value of a RPC function call.
Identifying the return value could vary from SUTs to SUTs, and from companies to companies.
So, we expose an interface to allow a customization of the identification (see Section~\ref{subsec:rpcextraction}).

Thus, with our RPC result analysis specification as shown in Figure~\ref{fig:rpcdto}, each result by a RPC function call would be constructed as an instance of \rpcresponse.
If there is an exception thrown, \rpcexceptioninfodto could be instantiated to describe info of exception in detail, such as exception class, message, type and category.
If a value is returned as defined, the value could be represented as a JSON object (if could) and an instance of \textit{ParamDto}, and the result could be further identified with \rpccustomizedresult.

\subsubsection{RPC Test Client}
This component mainly enables invocation of RPC function call with \rpcfunctiondto, analysis of the response or exception after the invocation, then outputting \rpcresponse.
With the \rpcfunctiondto, we could know what interface the action belongs to and what parameters are needed to construct, then the invocation is made based on the provided RPC client instance.
Result analysis is performed based on concepts we discussed in Section~\ref{subsubsec:resultanalysi}.
For instance, now we support extract name and message info from all exceptions which inherit from \texttt{java.lang.Exception}.
Its explicit type could be identified if it belongs to the Thrift framework, i.e., \texttt{org.apache.thrift.TException} can be found in the client library, the class of the thrown exception inherits from \texttt{TException}, then extract its super classes to recognize the exception category (e.g., \textit{APPLICATION}) under \rpcexceptioncategory and its \texttt{type} property to identify a type (e.g., \textit{APP\_INTERNAL\_ERROR}) under \rpcexceptiontype.
If the exception is not from Thrift framework, its explicit class would be extracted, then set it with \textit{CUSTOMIZED\_EXCEPTION} and \textit{UNEXPECTED\_EXCEPTION} based on whether the exception is a part of throw clause declared in the RPC function.
Note that the result analysis needs to be extended if one wants to support  other RPC frameworks, such as gRPC.
However, exceptions in the context of RPC domain have been formulated in our schema.
The additional work would be only technical details which we need to cope with, e.g., add additional types if they are not covered yet, then extract the specific info to identify the type.

More technical details on this implementation (e.g., how the parameters could be constructed for each data type, how to automatically recognize input parameters with customized info, and how to extract data and type from a Java object) can be found in our open source repository\textsuperscript{\ref{foot:link}}.

\subsubsection{Test Execution}
With our RPC handling support in the \textit{driver}, we enable tests to be executed during the search.
Then, with the JVM instrumentation provided by \evo, various SBST heuristics (e.g., code coverage, branch distance and SQL queries heuristics) can be returned after the test is executed (see \textit{JVM Instrumentation} $\rightarrow$ \textit{Fitness Function} in Figure~\ref{fig:approach}), additionally to the RPC function call execution results (i.e., \rpcresponse).
Regarding the authentication handling, dynamic tokens acquired via a login endpoint  can be regarded as an additional action which is needed to be invoked before the other RPC functions can be called.
This has been enabled automatically in our implementation.

\begin{table}
	\centering
	\caption{RPC testing targets with heuristics}
	\label{tab:rpc-targets}
	\begin{tabular}{p{0.01\linewidth}p{0.5\linewidth}p{0.12\linewidth}p{0.08\linewidth}p{0.08\linewidth}}
		\toprule
		\# & Execution Results & \textit{Handled} & \textit{Error} & \textit{isFault}\\
		\midrule
		1 & ER1: \textit{internal error} & 0.5 & 1 & Yes\\
		2 & ER2: \textit{user error} & 0.1 & 0.1 & -\\
		3 & ER4: \textit{other exception} & 0.5 & 1 & -\\
		4 & ER5-6: \textit{unexpected/declared}  & 0.5 & 1 & Yes\\
		5 & ER7: \textit{handled} & 1 & 0.5 & - \\
		\hline
		 &  & \textit{Success} & \textit{Fail} & \textit{isFault}\\
		\hline
		6 & \textit{success}  & 1 & 0.5 & -  \\
		7 & \textit{server error}  & 0.5 & 1 & Yes  \\
		8 & \textit{other error}  & 0.1 & 0.1 & -  \\
		\hline
		&  & \textit{NotNull} & \textit{Null} & \\
		\hline
		9 & a null response is returned & 0.5 & 1 & -  \\
		10 & a non-null response is returned & 1 & 0.5 & -  \\
		\hline
		&  & \textit{NotEmpty} & \textit{Empty} & \\
		\hline
		11 & an empty collection is returned & 0.5 & 1 & -  \\
		12 & a non-empty collection is returned & 1 & 0.5 & -  \\
		\bottomrule
	\end{tabular}
\end{table}

For white-box heuristics, we rely on the current state-of-the-art in white-box fuzzing of Web APIs given by \evo~\cite{Kim2022Rest,
zhang2022open}.
This includes adaptation of traditional SBST heuristics like the \emph{branch distance}~\cite{arcuri2019restful},
as well as advanced \emph{testability transformations}~\cite{arcuri2021tt} and \emph{SQL handling}~\cite{arcuri2020handling}.

In the context of testing RPC-based APIs, besides using SBST heuristics at code coverage level,
we propose additional novel testing targets (with their heuristics) on the responses of the RPC calls for guiding the test case generation, as shown in Table~\ref{tab:rpc-targets}.
Note that, with MIO, each testing target has a fitness value between 0.0 and 1.0, where a higher value is better.
A value with 1.0 means that the target is \textit{covered}, and any value more than 0.0 but less than 1.0 indicates that a testing target is \textit{reached} but not \textit{covered}.

For each RPC function, we create two testing targets \textit{Handled} and \textit{Error}, representing that the call is handled or in error, respectively, by the SUT.
Based on the execution results we reformulated in Section~\ref{sub:test-formulation}, we set a fitness value of \textit{Handled} and \textit{Error} testing targets as \#1-\#5 in Table~\ref{tab:rpc-targets}, after the call is executed.
For instance, if the execution result is identified as \textit{handled}, fitness values are set as 1.0 for \textit{Handled} and 0.5 for \textit{Error} (0.5 here represents the target is \textit{reached} but not \textit{covered}, which is heuristically better than not calling the method at all).
If any unexpected or declared exception is thrown, the fitness values are set as 0.5 for \textit{Handled} and 1.0 for \textit{Error}.
Since the exception type for the unexpected/declared exceptions is unclear, the execution would be further rewarded with a testing target for potential fault finding.
If the exception type could be further identified, the fitness values of \textit{Handled} and \textit{Error} would be handled as \#1-\#3.
Note that, for these three types of categorized exceptions, only \textit{internal error} is rewarded for potential fault finding.
Considering that the \textit{protocol error} typically refers to user errors, compared with other exceptions, it would be less important, then it is set with lower fitness values (i.e., 0.1) for both \textit{Handled} and \textit{Error}.
As \textit{transport error} (ER3) is usually due to issues in the testing environment (e.g., timeouts), then we do not reward such exception with any fitness values.

In addition, if the \textit{handled} results could be further categorized by user in terms of their business logic, we propose  two additional testing targets \textit{Success} and \textit{Fail}, representing whether the request succeeds or fails to be performed on the SUT.
Heuristics for handling the two targets regarding execution results are defined in \#6-\#8.
The strategy to decide the fitness values is similar with \textit{Handled} and \textit{Error} (e.g., \textit{server error} is rewarded with potential finding and \textit{other error} is recognized as less important) that aims at covering both \textit{Success} and \textit{Fail} of RPC function actions in terms of business logic.
Moreover, to maximize response coverage, we also propose another four testing targets by considering whether any null or non-null value is ever returned (i.e., \#9-\#10), and whether any empty or non-empty value is ever returned for collection datatypes (i.e., \#11-\#12).
Note that, although some of these fitness values do not provide much gradient for the search (e.g., only two values like $0.5$ and 1), they are still useful.
Test cases for reached but not covered targets (e.g., $0.5$) are kept in the archive of MIO, and will be still be sampled and mutated throughout the search.

\subsection{Search Operators}
\label{sub:searcoperators}

Our test reformulation enables its use in various search algorithms for supporting RPC-based API fuzzing.
In this work, we use MIO because it is the default in \evo, as it achieved the overall best results in an empirical study conducted by comparing it with various other algorithms~\cite{mio2017} on the fuzzing of RESTful APIs (recall Section~\ref{sub:evomaster}).
However, other search algorithms might be better on the problem of fuzzing RPC APIs.
But, without further empirical analyses, this is not something that can assessed for sure.
Due to the high cost of running this type of comparison experiments,
a comparison of different search algorithms for fuzzing RPC-based APIs is not in the scope of this paper.

MIO is an evolutionary algorithm inspired by (1+1) EA that uses two search operators, for sampling and mutation, respectively.
We employ the same strategies as \evo for RESTful API testing. 
The sampling is implemented to produce a valid test by selecting a sequence of one or more available actions at random.
Values of \textit{Gene}s in these tests are initialized at random, within the constraints if any (e.g., a \textit{ArrayGene} will have \textit{n} randomly generated  elements based on its min and max length). 
Authentication info, if any, is enabled with a given probability, i.e., 95\%, which is the default one used in \evo.
In addition, at the beginning of the sampling, we also prepare a set of adhoc tests which cover all available RPC function calls and all authentication combinations, i.e., each test has an action configured with and without authentication.
In other words, the structure of the first $k$ tests are not sampled at random, where $k=a\times n$, with $n$ being the number of functions in the RPC API and $a$ being the different authentication settings. 

Regarding the mutation operator, actions in a test can be added or removed for manipulating the structure of the test, given a certain probability.
To mutate values of \textit{Gene}s inside the tests, we employ the default value mutation in \evo, which has been integrated with \textit{taint analysis}~\cite{arcuri2021enhancing} and \textit{adaptive hypermutation}~\cite{zhang2021adaptive}.
How each gene is mutated depends on its type and constraints (if any), as previously discussed in Section~\ref{sub:test-formulation}.

Given a typical evolutionary algorithm with an individual representation having $n$ bits, then on average each gene would
be mutated with probability $1/n$.
However, the genes defined in \evo can have massive differences in terms of their genetic information.
For example, a boolean gene would represent only two possible values (for \texttt{true} and \texttt{false}), whereas an object gene
for a complex DTO could have hundreds of internal fields.
The search-engine of \evo can deal with genes of different \emph{weight}, and mutate the ones with more weight more often.
Furthermore, \textit{adaptive hypermutation}~\cite{zhang2021adaptive} enables having a higher mutation rate, and automatically detect which
genes have less (or none) impact on fitness, and automatically mutate them less often.

If the SUT interacts with a SQL database, genes to represent \texttt{INSERTION} operations will be automatically added to the tests, in the same way as done in \evo for RESTful APIs~\cite{arcuri2020handling}.

\subsection{Test Suite Output}
\label{sub:output}

In the same context of API testing,
we could re-use parts of \evo test writer to generate the SUT test scaffolding. 
For example, we use the same \texttt{initClass} for setting up the necessary testing environment (e.g., start SUT), \texttt{tearDown} for performing a cleanup after all tests are executed (e.g., shutdown SUT),  and \texttt{initTest} for resetting the state of the SUT for making test execution independent with each other.
To enable a more efficient test execution and fit industrial-scale API testing, we extended \texttt{initTest} with our smart database clean procedure, by considering the union of all accessed tables, and their linked tables, for all tests that are  generated.

Regarding handling of action execution and assertion generation,  
with \evo, tests are generated with \emph{RestAssured} to make HTTP calls toward the tested REST API. 
This is not applicable in the context of RPC testing.
Then, to support RPC-based API testing, we develop a \textit{Test Writer} that could handle instantiation of input parameters, RPC function call invocation (based on the RPC client library), and assertions on response objects with JUnit.
An example of generated tests can be found at this link\footnote{\url{https://github.com/anonymous-authorxyz/fuzzing-rpc/blob/main/example/src/em/EM_RPC_1_Test_others.java}}.

In our industrial case study, we found that some responses contain info such as time-stamps and random tokens, and they could change over time.
In order to avoid test failing due to such flakiness, we defined some general keywords (e.g., date, token, time) to highlight those cases.
If any keyword appear in either datatype, field name, or value with string type, the assertion would be commented out to avoid the test become flaky. 
We comment them out instead of removing them completely since it would be still interesting, for the users, to show what the response was originally. 

In addition, there might exist quite large responses in some API endpoints, especially when dealing with collections of data.
For example, in one SUT in our case study, a response contained 470 elements, and each element further contains data with list type, and 7579 assertions were generated for this response.
As such a large number of assertions would reduce the readability of the tests, we then developed a strategy to randomly select only $n$ (e.g., $n=2$) elements from the returned collections to generate assertions on in the tests.
More details on
the writer can be found in our open-source repository\textsuperscript{\ref{foot:link}}.

Generating this kind of tests has two main advantages.
First, as the generated tests are \emph{self-contained} (because they are able to start and stop the API directly without manual
intervention), they can be used for \emph{regression testing}.
Second, they help \emph{debugging} any found fault, as each generated test can be run independently, because they take care of initializing and reset the state of the API (e.g., SQL databases).
This feature was critical when analyzing the faults found during our empirical study.

\section{Empirical Study}
\label{sec:empirical}

\subsection{Research Questions}

In this paper, we conduct an empirical study to answer the following research questions: 
\begin{description}
	\item[{\bf RQ1}:] 
	 How does our white-box fuzzing perform compared with a baseline grey-box technique?
	\item[{\bf RQ2}:] 
	How does our novel approach perform in terms of code coverage?
	\item[{\bf RQ3}:] 
	Does our novel approach find real faults in industrial settings?
\end{description}

\subsection{Experiment Setup}

\begin{table}
	\centering
	\small
	\caption{Descriptive statistics of case studies}
	\label{tab:sut}
		
\begin{tabular}{lrrrrrr}
	\toprule 
	& \interface & \function & \service(\serviceU, \serviceD)  & \class & \loc$_f$ (\loc$_j$)   & \tableNumber\\
	\midrule 
	\rpcncs & 1           &        6 &  0  &      7 &      506 (\phantom{00}254)   &   0\\
	\rpcscs & 1           &        11  &  0  &    12  &     695 (\phantom{00}260)    &  0 \\
	\csA & 3           & 24         & 7 (\phantom{0}6, 1)    & 101      &    12559 (\phantom{0}4019)        & 6      \\
	\csB & 5           & 20         & 11 (\phantom{0}7, 4)   & 144      &     18987 (\phantom{0}1821)     &   17       \\
	\csC & 8           & 51         & 18 (14, 4)  & 339      &   45987 (18800)        & 156      \\
	\csD & 8           & 55         & 36 (27, 9)  & 868      &    116340 (20760)     & 50    \\
	\midrule

	\textit{Total}  & 26           & 167         & 72 (54, 18)  & 1471      &    195072 (45914)     & 229    \\
	\bottomrule 
\end{tabular}
\begin{spacing}{0.8}
	\raggedright \footnotesize 
	\interface represents the number of \textit{RPCInterface}s, 
	\function represents the number of available RPC functions, 
	\service represents the number of direct interacted external services (divided between \serviceU of upstream  and \serviceD downstream services), 
	\class is the number of Java class files, 
	\loc$_f$ is the number of lines of code in Files (\loc$_j$ is the number of lines of code reported by JaCoCo), 
	and \#Table is the number of SQL tables. 
\end{spacing}
\end{table}

To evaluate our approach (denoted as RPC-EVO), we carried out an empirical study with two artificial and four industrial RPC-based APIs selected by our industrial partner.
The industrial case studies are from a large-scale e-commerce platform (comprising hundreds of web services referred as microservices) developed by Meituan. 
Descriptive statistics of the case studies are summarized in Table~\ref{tab:sut}.  
\rpcncs and \rpcscs are re-implemented by us with Thrift, based on existing artificial RESTful APIs that have been used to assess the effectiveness of solving \textit{numeric} and \textit{string} problems~\cite{arcuri2019restful,arcuri2021enhancing,arcuri2020blackbox,zhang2021adaptive}.
\csA-\csD are from our industrial partner.
Each one of them is a part of large microservice architecture, where each API interacts with other services and a database.
\service (in Table~\ref{tab:sut}) shows the amount of external services that a SUT directly interacts with (see an example in Figure~\ref{fig:rpc-context}),
where \serviceU is the amount of its upstream services which the SUT depends on, and \serviceD is the amount of its downstream services which call the SUT.
The lines of code (\loc) numbers include everything, like comments, empty lines and import statements.
The actual lines with code (which results in \texttt{LINENUMBER} instructions in the compiled \texttt{.class} files) are calculated with the coverage tool JaCoCo (i.e., \loc$_j$).

Ideally, experiments should be carried on real industrial systems. 
However, we employed also two artificial case studies (which we open-sourced) to make at least parts of our experiments replicable\textsuperscript{\ref{foot:link}}, as of course we cannot share the code of the industrial systems.
In addition, to further demonstrate its adoption and performance in industrial settings, we also report preliminary results on {\totalindcs} further industrial APIs.
Note that the testing of those {\totalindcs} APIs was autonomously  performed by our industrial partner (e.g., prepare \evo drivers, without any researcher involved), as part of an internal evaluation to see whether/how to integrate \evo in their CI pipelines.

For the choice of baselines for comparisons,
regarding other tools in the literature,
to the best of our knowledge, there does not exist any other automated testing solution for RPC-based APIs that could be applied as a baseline in this study,
as discussed in more details in Section~\ref{sec:related}.
Therefore, we adapted our approach to be used by the Random Search Algorithm in \evo, which can be regarded as a grey-box technique (testing targets such as code coverage are still employed to evaluate tests to produce a final test suite as output at the end of the search).
This random search serves as a baseline to evaluate our approach in the context of white-box testing.
To be comparable,  the same search budget (i.e., 100 000 RPC function calls) are applied for all settings with these techniques.
In addition, to further evaluate the performance of our generated tests, we also compare them with existing tests in the industrial case studies.

Regarding the applied evaluation metrics,
in these comparisons we used three main metrics:
\emph{line coverage}, \emph{target coverage} and \emph{fault detection}.
Line coverage is measured with the code instrumentation of \evo, and it is based on the
\texttt{LINENUMBER} bytecode instructions in the business logic of the tested APIs (i.e., no third-party library).
Fault detection is based on the oracles defined in Section~\ref{sub:test-formulation}, i.e., it counts the
number of unique thrown exceptions related to server errors.
They are differentiated not only based on the entry point of the API (i.e., the actual method that is called remotely),
but also on the last executed statement in the business logic of the API (as each endpoint can fail for different reasons,
while executing different lines of the code).
The target coverage is an aggregated metric in \evo that considers \emph{all} its code coverage metrics
(e.g., besides covered lines, it also considers for example number of loaded classes, covered branches in jump instructions,
methods called without throwing exceptions,
and boolean methods returning \texttt{true} and \texttt{false} at least once),
 as well as the number of found faults, and black-box metrics on the API responses (recall Table~\ref{tab:rpc-targets}).

Considering the stochastic nature of the search algorithms, all experiments on each of the six main APIs were repeated 30 times, by following common guidelines in the literature~\cite{Hitchhiker14}.
\rpcncs and \rpcscs were executed on an HP Z6 G4 Workstation with Intel(R) Xeon(R) Gold 6240R CPU @2.40GHz 2.39GHz processor, 192G RAM, and 64-bit Windows 10.
The four industrial APIs were executed on the actual hardware pipelines of our industrial partner.
With these pipelines, all external services of the SUTs are up and running.
In an industrial testing environment, databases can be pre-loaded with lots of data (e.g., replicas of  the production database), for covering their specific  business logic.
The amount of such data can  be quite large, e.g., 256 024 data entries in \csC.
In an automated testing process, it is difficult to maintain such large data cost-effectively (e.g., clean and re-insert them back after each test execution) for ensuring that each test is executed with the same state of the SUT.
Therefore, we decided to use empty databases to conduct our experiments with the industrial APIs.

To get a better insight on the applicability of our novel techniques in real industrial contexts,
we also report on the use of \evo on the current testing pipelines at Meituan.
This included \totalindcs APIs, in which no researcher was involved in the running of these fuzzing sessions.
No comparison with \textit{grey box} technique was made here.
The preliminary results on these \totalindcs APIs are based only on 1 run, each one where \evo was run for 10 hours.

\subsection{RQ1: Comparison with Grey-box Technique}

\begin{table}
	\centering
	\small
	\caption{Pair comparisons between our approach (RPC-EVO) and Random with \#\textit{Targets} and \%\textit{Lines} on all SUTs.}
	\label{tab:comparison}
		\begin{tabular}{ l l r r r r r }\\ 
\toprule 
SUT & Metrics & RPC-EVO & Random & $\hat{A}_{12}$ & \emph{p}-value  & Relative \\ 
\midrule 
\emph{thrift-ncs}&{\#\textit{Targets}} & \textbf{542.2$\phantom{\%}$} & 376.3$\phantom{\%}$ & \textbf{1.00} & \textbf{$\le $0.001} & \textbf{+44.07\%} \\ 
&{\%\textit{Lines}$_e$} & \textbf{88.2\%} & 60.0\% & \textbf{1.00} & \textbf{$\le $0.001} & \textbf{+47.00\%} \\ 
\emph{thrift-scs}&{\#Targets} & \textbf{658.4$\phantom{\%}$} & 549.4$\phantom{\%}$ & \textbf{1.00} & \textbf{$\le $0.001} & \textbf{+19.85\%} \\ 
&{\%\textit{Lines}$_e$} & \textbf{71.9\%} & 59.6\% & \textbf{1.00} & \textbf{$\le $0.001} & \textbf{+20.59\%} \\ 
\emph{CS1}&{\#Targets} & \textbf{1953.3$\phantom{\%}$} & 1773.1$\phantom{\%}$ & \textbf{0.97} & \textbf{$\le $0.001} & \textbf{+10.16\%} \\ 
&{\%\textit{Lines}$_e$} & \textbf{25.6\%} & 23.1\% & \textbf{0.99} & \textbf{$\le $0.001} & \textbf{+11.15\%} \\ 
\emph{CS2}&{\#\textit{Targets} }& \textbf{3434.1$\phantom{\%}$} & 3099.1$\phantom{\%}$ & \textbf{0.96} & \textbf{$\le $0.001} & \textbf{+10.81\%} \\ 
&{\%\textit{Lines}$_e$} & \textbf{27.2\%} & 24.7\% & \textbf{0.94} & \textbf{$\le $0.001} & \textbf{+10.18\%} \\ 
\emph{CS3}&{\#\textit{Targets}} & \textbf{4453.4$\phantom{\%}$} & 4067.2$\phantom{\%}$ & \textbf{0.96} & \textbf{$\le $0.001} & \textbf{+9.50\%} \\ 
&{\%\textit{Lines}$_e$} & \textbf{15.4\%} & 14.0\% & \textbf{0.99} & \textbf{$\le $0.001} & \textbf{+9.78\%} \\ 
\emph{CS4}&{\#\textit{Targets}} & \textbf{6229.7$\phantom{\%}$} & 6046.8$\phantom{\%}$ & \textbf{0.80} & \textbf{$\le $0.001} & \textbf{+3.02\%} \\ 
&{\%\textit{Lines}$_e$} & \textbf{6.5\%} & 6.3\% & \textbf{0.91} & \textbf{$\le $0.001} & \textbf{+4.46\%} \\ 
\bottomrule 
\end{tabular} 

\end{table}

To answer RQ1, we applied our approach and the random search strategy
on all of the six case studies with the same search budget, i.e., 100 000 RPC calls.
The computation cost of two settings with at least 30 repetitions are $30.64$ hours for the two artificial APIs, and $129.12$ days for the four industrial APIs (maximum 18.27 hours and minimum 9.41 hours per run on the industrial APIs).
Note that the 30 repetitions are only applied in this empirical study for evaluating the approach.
When used by practitioners on their systems,  the approach can be run just once.

Table~\ref{tab:comparison} reports the results of target and line coverage, with comparison results on the two metrics using statistical analysis,
as recommended in~\cite{Hitchhiker14}.
In particular,
for pair comparisons we use the
Mann-Whitney-Wilcoxon U-test (see $p$-values) and Vargha-Delaney standardized effect size ($\hat{A}_{12}$).
The U-test is used with the standard $\alpha=0.05$ level (i.e., we claim statistical difference if a $p$-value is lower than $0.05$).
In this context, the $\hat{A}_{12}$ effect size provides a probability estimate that an algorithm produces better results than
the other compared one.
If two algorithms are equivalent, then $\hat{A}_{12}=0.5$.
A value like $\hat{A}_{12}=1$ means that, in every single run, the first algorithm always gave better results.
Note that $\hat{A}_{12}$ effect size only computes \emph{how often} an algorithm gives better results, but not by \emph{how much}.

Given these results, our approach demonstrates significantly better results than the baseline technique, with a low \textit{p-}value (i.e., $\le $ 0.001) and a high effect size (i.e., $\hat{A}_{12}$ > 0.80) on all of the six case studies with the two metrics.

\begin{figure}
	\begin{subfigure}{.33\textwidth}
		\includegraphics[width=\linewidth]{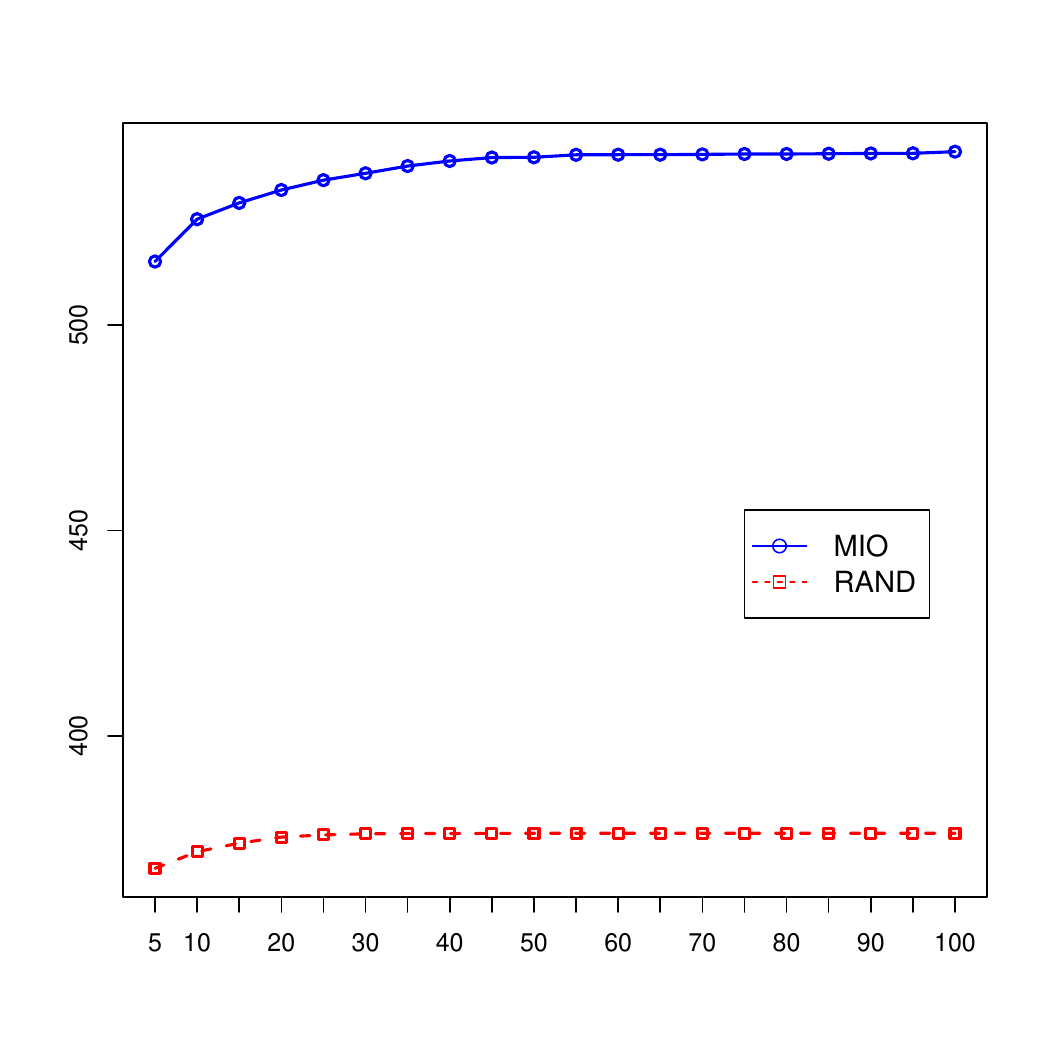}
		\caption{\rpcncs}
		\label{fig:ncs}
	\end{subfigure}\hfill
	\begin{subfigure}{.33\textwidth}
		\includegraphics[width=\linewidth]{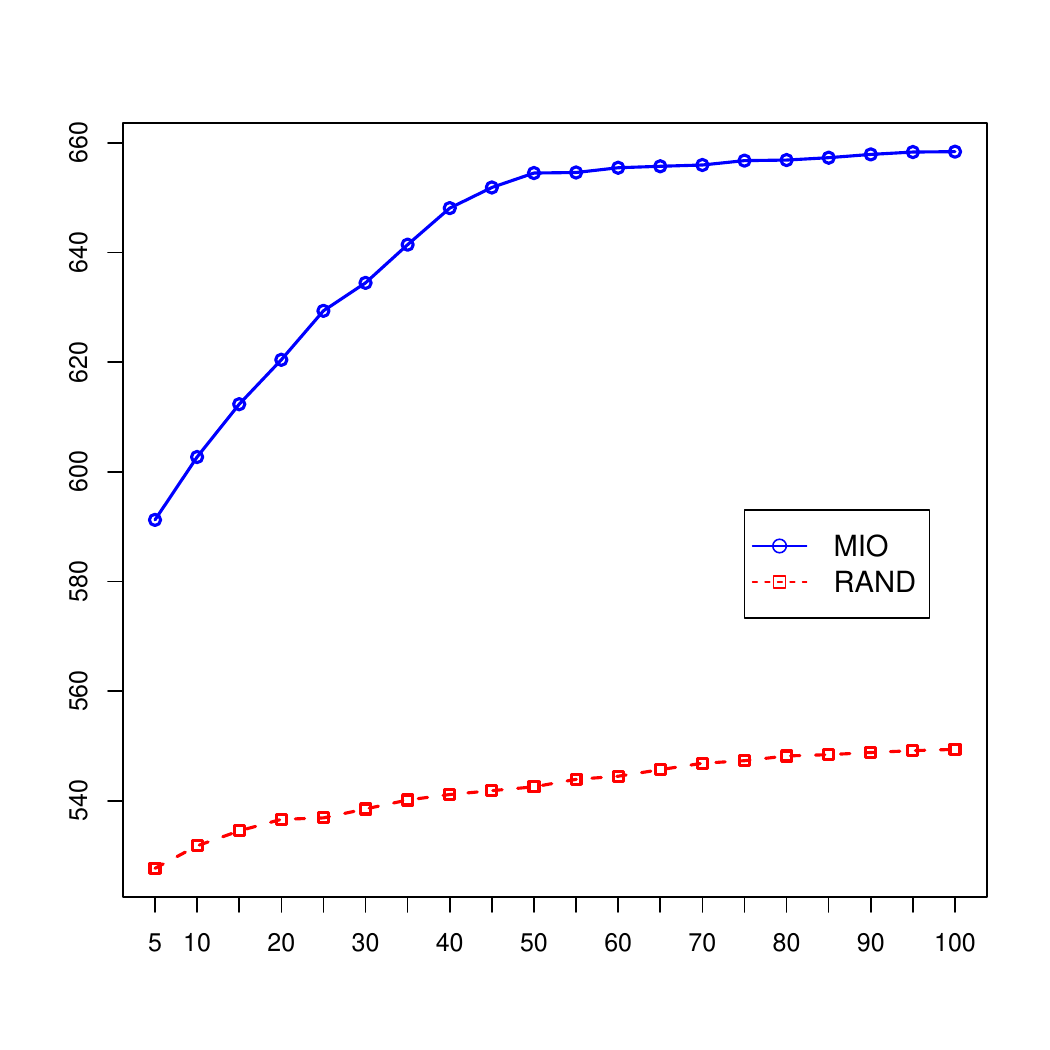}
		\caption{\rpcscs}
		\label{fig:scs}
	\end{subfigure}\hfill
	\begin{subfigure}{.33\textwidth}
		\includegraphics[width=\linewidth]{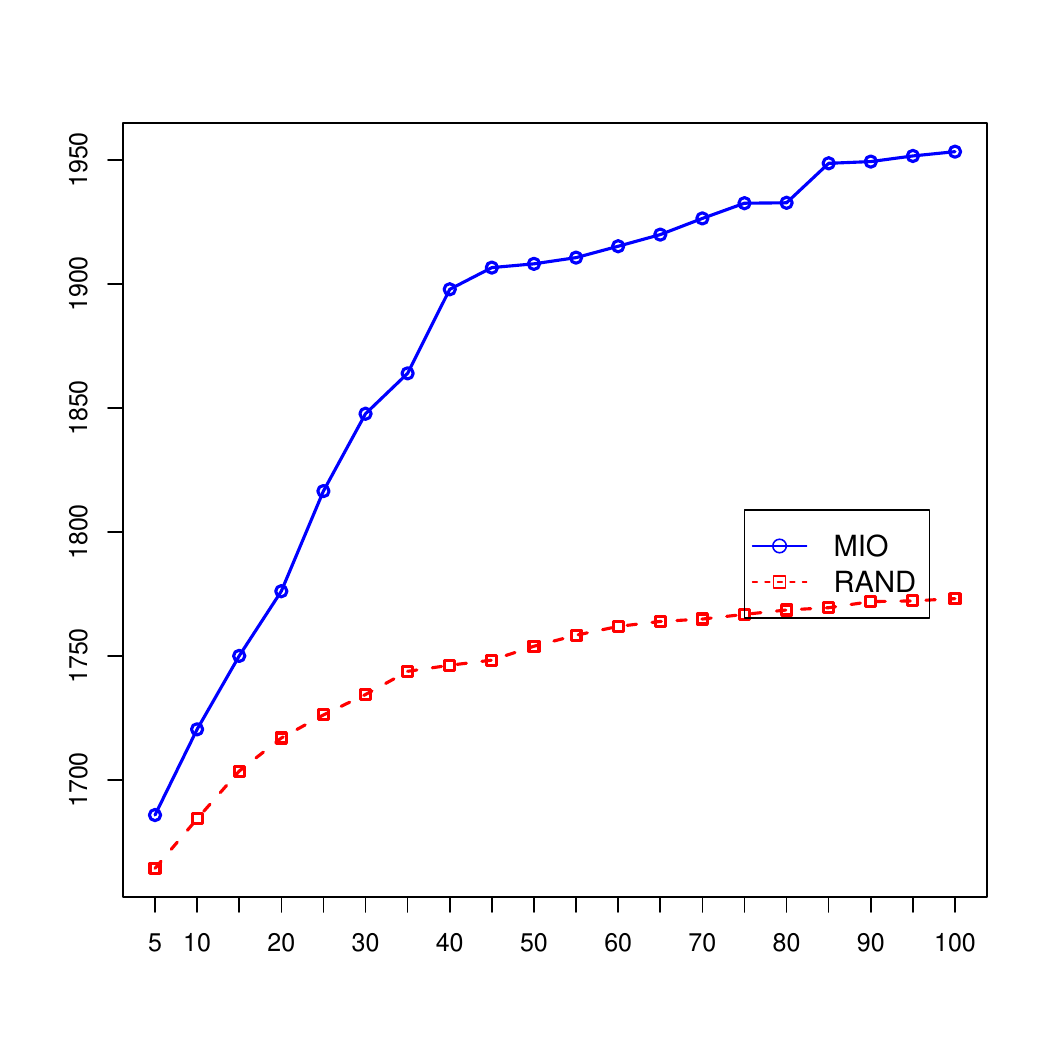}
		\caption{\csA}
		\label{fig:cs1}
	\end{subfigure}
	\begin{subfigure}{.33\textwidth}
		\includegraphics[width=\linewidth]{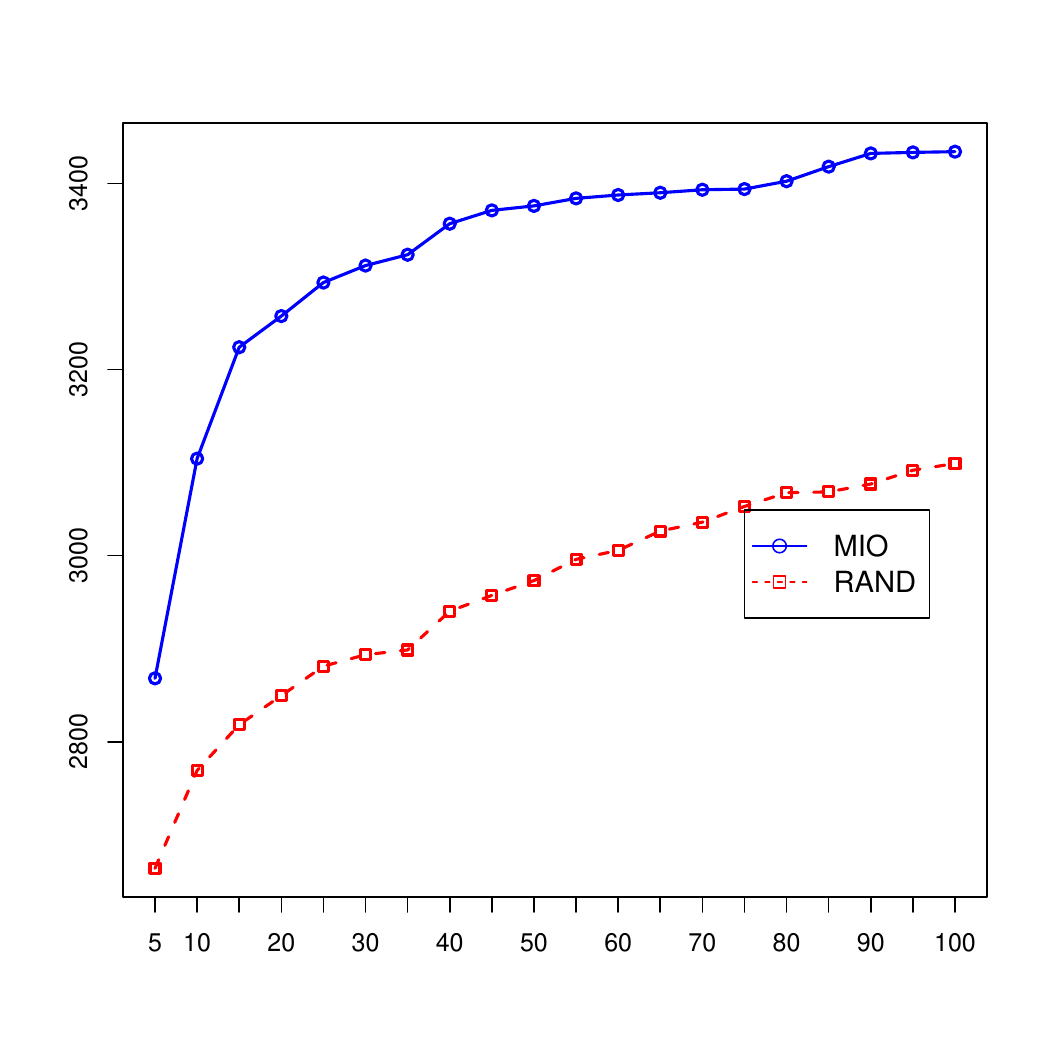}
		\caption{\csB}
		\label{fig:cs2}
	\end{subfigure}\hfill
	\begin{subfigure}{.33\textwidth}
		\includegraphics[width=\linewidth]{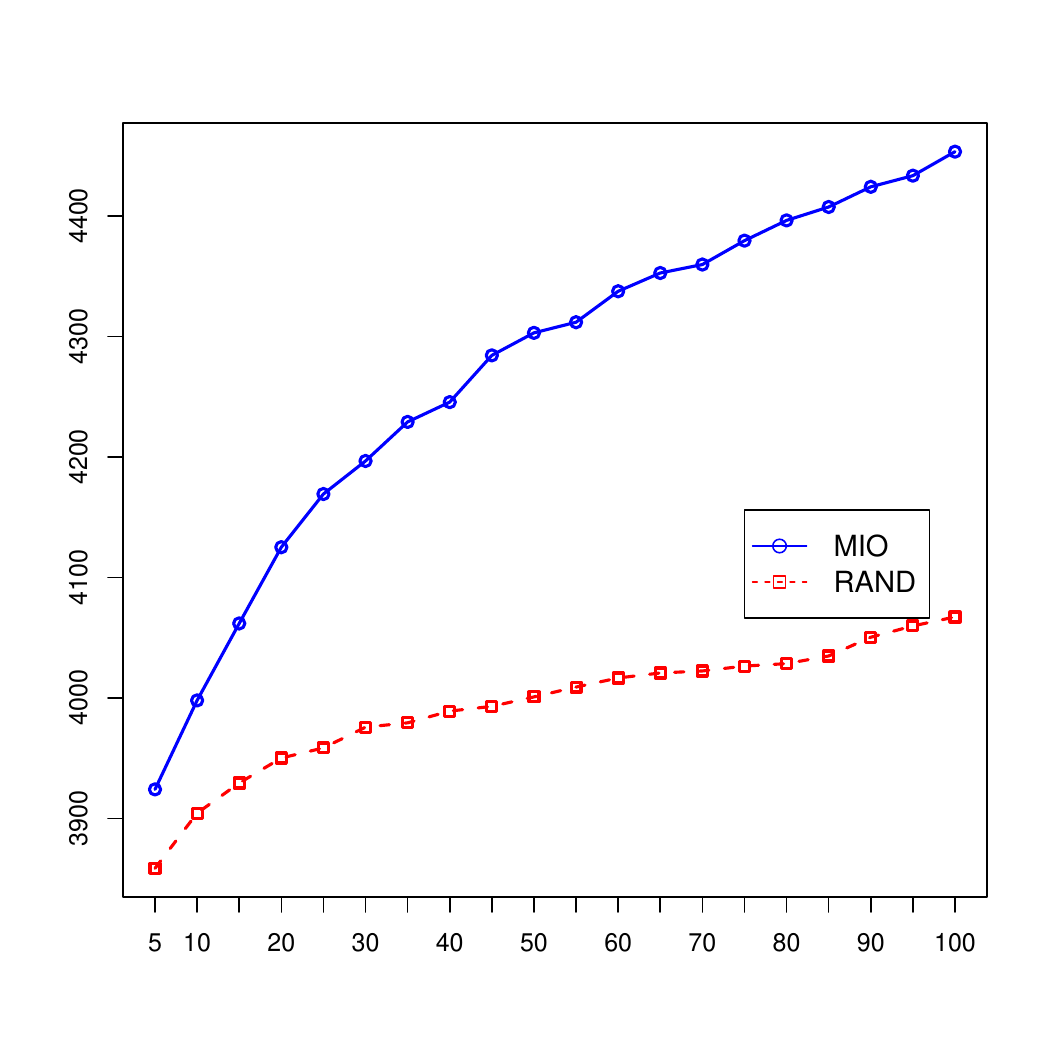}
		\caption{\csC}
		\label{fig:cs3}
	\end{subfigure}\hfill
	\begin{subfigure}{.33\textwidth}
		\includegraphics[width=\linewidth]{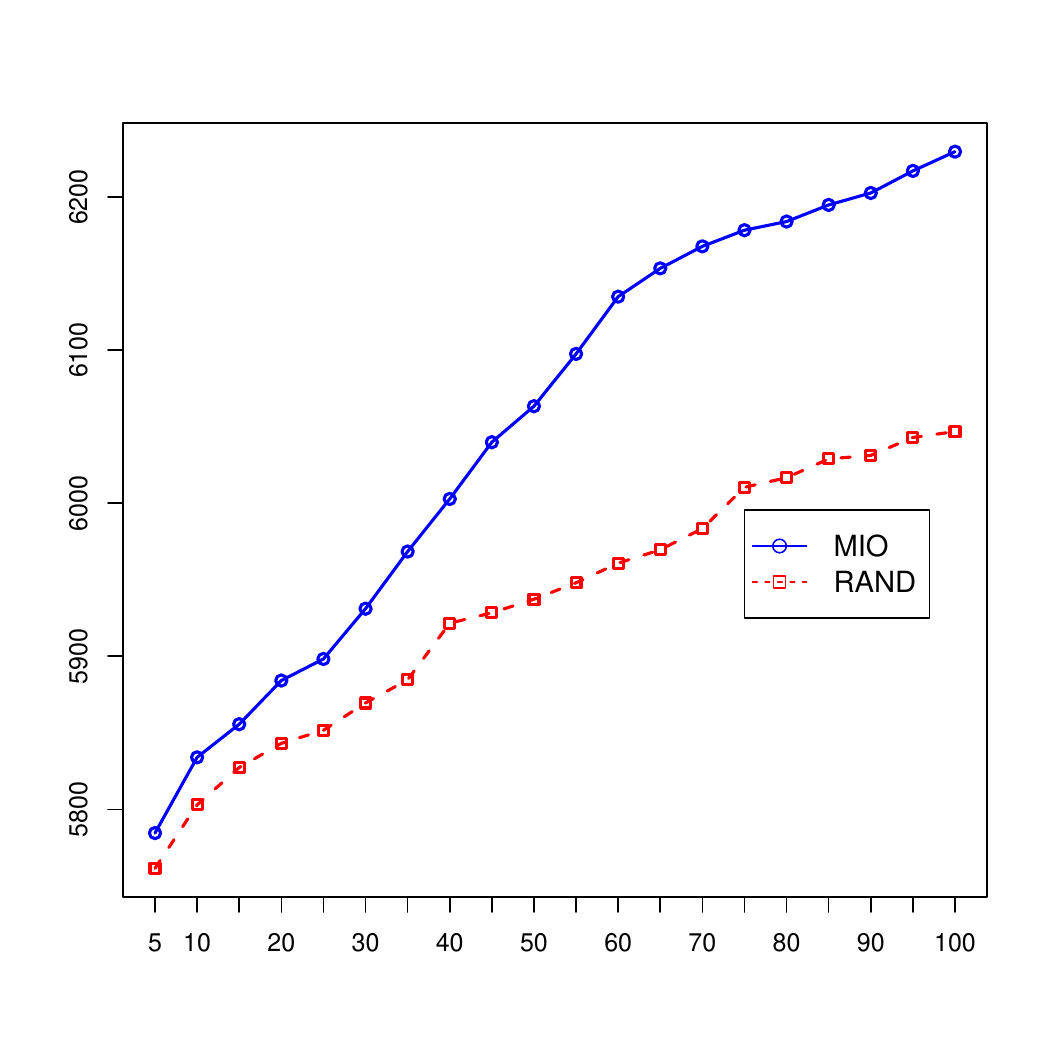}
		\caption{\csD}
		\label{fig:cs4}
	\end{subfigure}
	\caption{At every 5\% of the used budget (x-axis), average covered targets (y-axis) achieved by RPC-EVO and Random. }
	\label{fig:targets}
\end{figure}

In addition, Figure~\ref{fig:targets} plots the average covered targets over time (i.e., at every 5\% of the used budget) for two techniques on each case study.
Based on these line plots, 
our approach clearly outperforms Random by a large margin throughout the whole process of the search, and the results are consistent on all of the case studies.  
This further demonstrates the effectiveness of our white-box techniques in both artificial and industrial settings.
 
\begin{result}
	RQ1: Based on the target and line coverage results, our approach significantly outperforms random search on all of the six case studies. The relative improvements are up to 47\% on the artificial case studies and 11.15\% on the industrial case studies. 
\end{result}

\subsection{RQ2: Results of Code Coverage}

\subsubsection{Artificial APIs}
Based on coverage (i.e., \%\textit{Lines}$_e$) reported in Table~\ref{tab:rpc-targets},
on the two artificial case studies representing numeric and string testing problems, our approach achieves  high line coverage (i.e., 88.2\% on \rpcscs and 71.9\% on \rpcscs) when using  100\textit{k}  calls as budget.   
This high code coverage could demonstrate that RPC-EVO effectively enables the white-box fuzzing for RPC-based APIs, i.e.,  based on the white-box heuristics, effectively optimize the inputs of extracted/reformulated RPC function calls.

\begin{result}
	RQ2.1: Our approach achieves  high line coverage on the two artificial case studies, demonstrating its effectiveness in enabling white-box fuzzing of RPC APIs. 
\end{result}

\subsubsection{Industrial APIs}
\label{subsubsec:rq2_industrial_apis}
Regarding the four industrial case studies, as a fully automated solution, our approach achieved useful (for our industrial partner) coverage on \csA and \csB (more than 25\%), but limited coverage on \csC and \csD (especially \csD).
The results are also related to the complexity of these SUTs (as shown in Table~\ref{tab:sut}), given the  same limited search budget.
For example, based on \textit{\#LoC} values, \csC and \csD are much larger (and likely more difficult to fully cover) than \csA and \csB, where \csD has more than 2.5 times \textit{\#LoC} than \csC.
In addition, based on the line plots in Figure~\ref{fig:targets}, the slope of the lines in \csA-\csD is greater than \rpcncs and \rpcscs, especially for \csC and \csD.
This indicates that more targets  would likely be covered if more budget is used, i.e., if the fuzzers were run for longer, like 24 or 48 hours. 
However, without actual experiments, it is not possible to be completely sure.

\textbf{Comparison with existing tests.} 
To study the performance on code coverage in industrial settings, we compare our generated tests with existing tests.
The analysis was conducted with three groups of tests: 
\begin{itemize}
	\item $W$: a test suite generated by RPC-EVO that achieves the worst result  out of the 30 repetitions; 
	\item $B$: a test suite generated by RPC-EVO that achieves the best result; 
	\item $E$: a set of existing tests.
\end{itemize}
The code coverage was collected by executing the generated tests and existing tests on the SUTs with Intellij~\cite{intellijCov}.

\begin{table}
	\centering
	\small
	\caption{Numbers of generated test cases in the worst run (RPC-EVO$_w$) and the best run (RPC-EVO$_b$) of our approach out of the 30 repetitions, compared to the number of existing tests in the industrial APIs.}
	\label{tab:numtests}
		\begin{tabular}{lrrrr}
	\toprule 
	\textit{SUT} & RPC-EVO$_w$ & RPC-EVO$_b$ & Existing (Manual, Replay) \\
	\midrule
	\csA & 80 & 75 & 12 (\phantom{0}0, 12)\\
	\csB & 89 & 71 & 5 (\phantom{0}5, \phantom{0}0)\\
	\csC & 186 & 156 & 27 (\phantom{0}1, 26) \\
	\csD & 236 & 232 & 46 (12, 34) \\
	
	\bottomrule 
\end{tabular}
\end{table}

In Table~\ref{tab:numtests}, we report the numbers of tests in these groups for each of \csA--\csD.
Regarding the existing tests in the industrial setting of our partner,
those are the actual tests currently used at Meituan to find regression faults in these APIs.
These tests were prepared by engineers and testers at Meituan, completely independently from our experiments.
There exist two types of these tests.
One type is manually written automated tests (e.g., JUnit), and the other is with a replay of manual testing (using a custom testing tool).
The manual testing would be driven from the user-side, as the example shown in Figure~\ref{fig:rpc-context}.
For instance, a tester
at Meituan would
perform a real business scenario as a user, by directly interacting with an app on their mobile phones.
Then, with the requests from the user, it would result in various RPC communications among the services.
Those communications are recorded (such as what calls are invoked) along with the states of the connected external services.
With an industrial testing tool, such records are performed as a replay (such as re-execute the calls) on the SUTs for conducting manual regression testing.
In order to collect code coverage for the record for the comparisons in this paper,  we converted the records as JUnit tests by extracting the calls and their inputs (but a setup of the states of external services with the replay tool cannot be transferred into the JUnit tests).

\begin{table}
	\centering
	\small
	\caption{Results of
		line coverage achieved by the worst run and the best run of our approach (denoted as $W$ and $B$ respectively) and existing tests (denoted as $E$). ``Total'' represents the union coverage achieved by RPC-EVO and existing tests, and ``Uncovered'' represents the  coverage achieved by the existing tests but not by our approach. }
	\label{tab:comp-exisiting}
	\begin{tabular}{l rrrr}
	\toprule 
	& RPC-EVO (\%)  & Existing (\%) & Total (\%) & Uncovered (\%) \\
	\textit{SUT} & [$W$, $B$] & $E$ & [$W \cup E$, $B\cup E$] & [$ E \setminus W$, $E \setminus B$] \\
	\midrule
	\csA & [22.50, 26.81] & 14.29 & [22.95, 27.85] & [0.45, 1.04] \\
	\csB & [25.46, 26.02] & 16.28 & [25.46, 26.40] & [0.00, 0.37] \\
	\csC & [13.37, 15.89] & 5.29 & [13.46, 15.92] & [0.09, 0.03] \\
	\csD & [\phantom{0}\underline{8.31}, \phantom{0}9.15] & 8.31 & [10.22, 10.90] & [1.91, 1.75] \\
	\midrule
	\textit{Avg.} & [17.41, 19.47] & 11.04 & [18.02, 20.27] & [0.61, 0.80] \\
	\bottomrule 
\end{tabular}
\end{table}

Results of line coverage reported with Intellij achieved by the three groups for \csA--\csD are reported in Table~\ref{tab:comp-exisiting}.
Note that the line coverage might be slightly different with the results in Table~\ref{tab:comparison} which are reported with \evo bytecode instrumentation.
However, the comparison is always performed with results obtained from the same coverage runner.

In this table, we also provide a union of code coverage achieved by RPC-EVO and existing tests (see \textit{Total\%}), and a code coverage achieved by the existing tests but not RPC-EVO (see \textit{Uncovered\%}).
Then we observed that
\begin{itemize}
	\item
	First, by checking \textit{Total\%} with existing tests $E$ in industrial APIs, on all SUTs,
	RPC-EVO (i.e., both the worst and the best test suites) can attribute to additional code coverage compared with the existing ones;
	\item
	By comparing RPC-EVO with $E$, on all of the four industrial case studies, the code coverage by RPC-EVO (i.e., the worst and the best) are clearly greater than the existing ones, except \csD which achieves the equivalent results (i.e., \underline{8.31}\%);
	\item
	Regarding \textit{Uncovered}, the percentage is minor (i.e., the max is 1.91\%).
	This indicates that
	RPC-EVO is capable of covering  most of the code achieved by the existing tests, i.e., above 77.02\% (=$1-1.91/8.31$), up to 100\% of the code covered by $E$;
	\item
	One interesting observation here is that the selected worst run perform better in covering lines achieved by existing tests on \csA and \csB than the best run. 
	This might further reveal various promising regions of search space in industrial problems.
\end{itemize}
Based on the observation, we can conclude:

\begin{result}
	RQ2.2: Compared with existing tests, RPC-EVO is capable of contributing additional code coverage and demonstrates  clear better results. In addition, RPC-EVO could cover above 77.02\% line coverage achieved by the existing tests.
\end{result}

\textbf{In-depth analysis of the coverage reports.}
To further study why higher coverage was not achieved, 
we  performed a manual analysis on the code coverage reports generated by the best test suite (i.e., $B$) and the source code of these APIs.
For \csD, we found that 10 RPC functions out of the 55 functions are not accessible with the given client library.
By checking with our industrial partner, they think these problems are due to some issues in their testing environment (e.g., which uses a service discovery mechanism and load balancers) that were found as well in executing existing tests.
These problems are currently under investigation.
This might be a reason for the least line coverage 
achieved by our approach on \csD.

Based on the coverage reports on the four APIs, we found that our approach achieved limited code coverage on the code which is related to database handling and communications with external services. 
Regarding the database handling code, most of it is automatically generated with an in-house framework for facilitating various manipulations on the database, e.g., to perform a query with various conditions.  
As discussed with our industrial partner, usually, not all of the generated manipulation code would be used in implementing their business logic.
However, they are still generating and keep such code in case of future use.
Therefore, a lot of this code is infeasible to be covered with any system test. 

Regarding the code related to the communications  with the external services, all these services were up and running in the test environment of our industrial partner.
Since these external services are not mocked nor their code is instrumented with our SBST heuristics, i.e., not being part of our testing process, then we mostly fail to get different responses  with our automatically generated inputs that maximize the code coverage in the SUT (e.g., all the code used to read and act upon the responses given by these external services).
How to deal with external services is a major research challenge which applies to all kinds of web services.

Another main issue that we found is related to input validation.
In our approach, we have handled all the constraints specified as parts of interface definitions, but there also exist further restricted checks on the inputs.
The inputs could be restricted in the internal business logic of the SUT, e.g., an input parameter \textit{x} could be validated with an external service regarding whether it exists: then, if it exists, it could further query another service for the related data with \textit{x}.
With our current heuristics, such valid inputs could be rarely generated.
Moreover, the needed inputs are often complex.
For instance, we noticed that, in a generated test, there are 2024 lines for instancing a single input DTO.
As we checked, the length for all lists in that instance is less than 5.
Then, we further checked the implementation of the DTO, which it contains 25 fields, and the fields could be other DTOs or  lists of DTOs. 
Such very large DTO would lead to additional difficulty to generate valid inputs, e.g., if any element (e.g., in a collection) violates any constraint, then the whole DTO would be considered as invalid, and fail the input validation.

Not that there could be more issues at play here that could explain these results.
This could include the complexity of the source code, and/or possible side effects of existing search algorithms such as MIO on this problem domain.
Without further in-depth analyses, it is currently not possible to pin-point the main culprit for this low coverage.
Regardless, the identified issues will need to be addressed, paving the road ahead for further research on improving the fuzzing of RPC APIs.
Once fixed, re-running these experiments will be needed to identify if there are still any major issues impacting the achieved code coverage.

\begin{result}
	RQ2.3: 
	Our approach achieves
	useful coverage (26.81\% and 26.02\%) on two out of the four industrial case studies, and limited coverage (15.89\% and 9.15\%) on the other two larger industrial case studies.
	Based on a manual analysis on code coverage and the source code, we found that the main issues are related to the communications with external services and to generate inputs for complex DTO with various constraints. 
\end{result}

\begin{table}
	\centering
	\small
	\caption{Results of line coverage of 50 industrial APIs achieved by RPC-EVO with 1 run using 10 hours budget.}
	\label{tab:ind30}

			\begin{tabular}{ l r r r r | l r r r r}\\ 
\toprule 
\# & \#\textit{LoC}$_j$  & \#\textit{Targets} & \%\textit{Lines} & \#\textit{Faults} & \# & \#\textit{LoC}$_j$    & \#\textit{Targets} & \%\textit{Lines}& \#\textit{Faults} \\ 
\midrule 
\#01 & 19867 & 26439 & 15.44 & 59 & \#02 & 54457 & 34141 & 15.02 & 90\\ 
\#03 & 39308 & 11459 & 14.53 & 82 & \#04 & 31663 & 27485 & 34.65 & 129\\ 
\#05 & 34872 & 23363 & 28.44 & 71 & \#06 & 27859 & 14071 & 19.38 & 172\\ 
\#07 & 38179 & 15960 & 17.96 & 168 & \#08 & 39075 & 25665 & 27.41 & 295\\ 
\#09 & 58232 & 81279 & 17.15 & 292 & \#10 & 28814 & 14556 & 21.93 & 146\\ 
\#11 & 63108 & 81200 & 9.37 & 371 & \#12 & 30556 & 40379 & 16.03 & 265\\ 
\#13 & 30954 & 29830 & 6.11 & 252 & \#14 & 40644 & 16720 & 14.1 & 179\\ 
\#15 & 7314 & 11280 & 15.72 & 66 & \#16 & 34369 & 48320 & 16.51 & 315\\ 
\#17 & 80929 & 72592 & 11.95 & 354 & \#18 & 38914 & 22582 & 15.83 & 160\\ 
\#19 & 16880 & 13726 & 21.59 & 56 & \#20 & 6597 & 11284 & 16.65 & 141\\ 
\#21 & 6882 & 2294 & \underline{4.3} & \underline{4} & \#22 & 2019 & 5478 & 20.97 & 16\\ 
\#23 & 53565 & 35282 & 11.49 & 99 & \#24 & 28604 & 102260 & 10.42 & \textbf{2250}\\ 
\#25 & 15047 & 27929 & 9.44 & 82 & \#26 & 58578 & 54496 & 15.01 & 164\\ 
\#27 & 52878 & 13921 & 12.64 & 127 & \#28 & 8548 & 4735 & 19.85 & 35\\ 
\#29 & 89303 & 18350 & 12.64 & 233 & \#30 & 9782 & 1442 & 8.17 & 19\\ 
\#31 & 10054 & 5273 & 22.3 & 67 & \#32 & 23864 & 5445 & 9.11 & 83\\ 
\#33 & 21055 & 10625 & 27.48 & 109 & \#34 & 19191 & 4709 & 12.25 & 41\\ 
\#35 & 6602 & 6161 & 27.72 & 40 & \#36 & 10676 & 6059 & 15.29 & 85\\ 
\#37 & 16534 & 24051 & 9.02 & 95 & \#38 & 23349 & 11695 & 24.84 & 57\\ 
\#39 & 7006 & 6574 & 17.8 & 14 & \#40 & 8096 & 2478 & 8.95 & 16\\ 
\#41 & 12081 & 3840 & 14.41 & 35 & \#42 & 20769 & 7744 & 15.03 & 55\\ 
\#43 & 41450 & 13774 & 12.72 & 74 & \#44 & 2217 & 8533 & 15.99 & 32\\ 
\#45 & 58439 & 20297 & 19.3 & 148 & \#46 & 10230 & 6059 & 10.5 & 13\\ 
\#47 & 56242 & 20230 & 17.69 & 235 & \#48 & 9691 & 33559 & 23.97 & 74\\ 
\#49 & 3301 & 4594 & \textbf{59.16} & 63 & \#50 & 35401 & 19609 & 30.4 & 349\\ 
\hline 
\multicolumn{2}{l}{\textbf{\#\textit{LoC}$_j$}} & \multicolumn{8}{r}{\textbf{Sum}: 1444045; \textbf{Avg}: 28880.90; \textbf{Max}: 89303; \textbf{Min}: 2019} \\ 
\hline 
\multicolumn{2}{l}{\textbf{\%\textit{Lines}}} & \multicolumn{8}{r}{\textbf{Avg}: 17.49; \textbf{Max}: 59.16; \textbf{Min}: 4.3} \\ 
\multicolumn{10}{r}{\textbf{60$\sim$50\%}: 1; \textbf{50$\sim$40\%}: 0; \textbf{30$\sim$40\%}: 2; \textbf{20$\sim$30\%}: 10; \textbf{10$\sim$20\%}: 29; \textbf{4$\sim$10\%}: \phantom{1}8} \\ 
\hline 
\multicolumn{2}{l}{\textbf{\#\textit{Faults}}} & \multicolumn{8}{r}{\textbf{Sum}: 8377; \textbf{Avg}: 167.54; \textbf{Max}: 2250; \textbf{Min}: 4} \\ 
\bottomrule 
\end{tabular} 

	\begin{spacing}{0.8}
		\raggedright \footnotesize
			\# is an index of industrial SUTs; \loc$_j$  is the number of lines of code reported by JaCoCo; \textit{\#Targets} is the number of targets covered by our approach that is composed of lines, branches and potential faults; \textit{\%Lines} is the line coverage achieved by our approach; \textit{\#Faults} is the number of potential faults identified by our approach.
			
	\end{spacing}
\end{table}

\textbf{Additional analysis with code coverage collected by our industrial partner.} 
In Table~\ref{tab:ind30}, we report the preliminary results of target coverage and line coverage in {\totalindcs} industrial APIs achieved by RPC-EVO with 1 run using 10 hours search budget.
Note that all these {\totalindcs} industrial APIs plus \csA--\csD are parts of one single microservice architecture, with hundreds of web APIs.
The 10-hour budget was decided by our industrial partner by considering their application context and time cost per run in this experiment.

Based on the results, on {\totalindcs} industrial APIs with {\totalIndLoc} lines of code (\loc$_j$) in total and {\avgIndLoc} lines of code on average {(ranging from \minIndLoc to \maxIndLoc)}, RPC-EVO achieves 30\%$\sim$60\% line coverage on 3 SUTs,  20\%$\sim$30\% line coverage on 10 SUTs,  10\%$\sim$20\% line coverage on 29 SUTs, and 4\%$\sim$10\% line coverage on 8 SUTs.
These code coverage results are consistent with the ones reported in Table~\ref{tab:comparison} for the other 4 APIs we analyzed in more details.

\begin{result}
	RQ2.4: RPC-EVO has been successfully applied in white-box fuzzing {\totalindcs} industrial RPC-based APIs in practice by our industrial partner. 
	With 10-hour search budget, results show that our approach is capable of achieving on average \avgIndLineCoverage (up to \maxIndLineCoverage) line coverage.
\end{result}

\subsection{RQ3: Results of Fault Detection}
\label{subs:results-faults}

\begin{table}
	\centering
	\small
	\caption{Results of the potential distinct faults automatically reported by our approach with 30 runs for each industrial SUT. We report as well the number of real faults manually identified and confirmed with the industrial partner.}
	\label{tab:faults}
		\begin{tabular}{ l r |r r r r}\\ 
\toprule 
& Potential Faults & \multicolumn{4}{c}{Real Faults} \\ 
SUT & Avg.[Min, Max]  & L1  & L2 &L3 & \textit{Total}\\ 
\hline
\emph{CS1}&40.2 [40, 41]&  17 & 0 & 5 & 22\\ 
\emph{CS2}&21.8 [21, 26]&  3 & 1 & 15 & 19\\ 
\emph{CS3}&51.6 [51, 55]&  1 & 0 & 29 & 30\\ 
\emph{CS4}&91.8 [74, 111]& 3 & 16& 36 & 55\\ 
\hline
\textit{Total} & & \textbf{24} & \textbf{17} & 85 & 126 \\
\bottomrule 
\end{tabular} 

	\begin{spacing}{0.8}
		\raggedright \footnotesize \textit{L1}: faults  that will be fixed; \textit{L2}: faults are needed to be fixed but less important; \textit{L3}: faults that are tolerable, and likely no need to fix
	\end{spacing}
\end{table}

To assess the fault detection capabilities of our novel approach, we performed a detailed analysis on the identified faults with our industrial partner,
as researchers and industrial practitioners might have different views on the severity and importance of the found faults.
The manual analysis is based on the test suites that achieved the best code coverage (out of the 30 runs) for each of the four industrial APIs we analyzed in details.
We applied such selection  due to the time constraints of manually checking all the generated test suites  in all the 30 repetitions. 
With this selection, the amount of tests to be reviewed are 534 as  RPC-EVO$_b$ shown  in Table~\ref{tab:numtests}.
With these tests, faults are identified based on (1) any exception thrown in the calls; (2) service error represented by assertions on the responses; (3) failed tests when executing them on the SUT (mainly due to flaky assertions); and (4) whether responses are expected based on the given inputs.
The review was firstly conducted by the first author, then an employee of our industrial partner (a QA Manager who has 8 years of testing experience in industry) performed the same kind of analysis on these tests.
At the end, a meeting was held to discuss and confirm the final results reported in Table~\ref{tab:faults}.

As shown in Table~\ref{tab:faults}, in total 126 real unique faults were found with the selected test suites on the four industrial APIs.
The faults could be further classified into three levels, i.e., \textit{L1}, \textit{L2} and \textit{L3}, based on willingness of our industrial partner to fix these faults.
\textit{L1} represents the number of faults that are serious enough that should be fixed. 
These faults are related to mistakes in the code implementation, errors in handling databases, errors in transaction processing and potential risky errors in returning a misleading response.
At the time of writing this paper, the identified faults have all been confirmed and fixed. 
\textit{L2} is the number of faults that should be fixed but are less critical.
Most of the faults at \textit{L2} are related to the implementation of input validation and external service response handling when exceptions are thrown.
In their context, it is better to properly handle exceptions within the SUT, as such thrown exceptions might lead to further problems in the services that depend on the tested application.
\textit{L3} is the number of minor faults that are tolerable, and most likely our industrial partner will not fix them.
These faults are mainly due to input validation throwing exceptions such as \texttt{NullPointerException}, \texttt{IndexOutOfBoundsException} and \texttt{java.text.ParseException}.
However, if the exceptions are caught and handled within the SUT, they consider that such faults are tolerable.

In Table~\ref{tab:faults}, we also report the number of potential faults automatically reported by our approach.
As expected, the number of real faults we manually identified is less than the potential ones.
This is mainly due to (1) problems in the test environment (e.g., some external services might not had been up and running when the experiments were carried out); (2) data preparation in databases (e.g., an empty database might lead to some problems that would never happen in production); (3) communications over the network (e.g., connection timeouts); (4) client problems 
(e.g., some remote functions fail for some configuration issues when we ran the experiments for \csD).
However, with the generated tests, our employed automated oracles could identify most of the real faults, except the errors related to returning a misleading/unexpected response (as this requires the users to manually check the content of these responses, as no formal specification is available).

Regarding the further experiments with 1 run on a further \totalindcs APIs,
RPC-EVO identified in total \totalIndfaults  and on average \avgIndfaults potential faults in these \totalindcs industrial APIs as shown in Table~\ref{tab:ind30}.
For the industrial API  \#24, the number of detected faults is significantly higher than for the other APIs (i.e., \maxIndfaults faults).
By performing a further investigation on this API, we found that the API is for handling authentication, and its functions are invoked by many other services in the microservices.
Thus, a request to this API requires to link with a valid account and be specified with valid data, i.e., the account should have an access to the data, and the data should be accessible and satisfy corresponding business features for the account.
Any request volatilizing such constraints would throw exceptions under the current implementation.
With a 10 hour search budget, tests generated by our approach led to such unexpected exceptions thrown from 2,091 different locations in the  code.
This might explain why such a high number of potential faults were detected in this API.

At this point in time, we do not know yet how many of these detected \totalIndfaults faults are critical, and must be fixed as soon as possible.
This is currently under evaluation by the engineers and testers at Meituan.
Going through and debugging thousands of potential faults is a time consuming task.

\begin{result}
	RQ3: With an in-depth analysis of the generated tests with our industrial partner, we confirm that our approach was capable of finding 41 actual real faults that have now been fixed.
	Another \totalIndfaults potential faults are currently under investigation.
\end{result}

\section{Lessons Learned}
\label{sec:lessons}

\textbf{Automated testing requires a reset of the SUT, however, it is challenging to reset the state of a real industrial API.}
To enable the generated tests to be used for regression testing, and to properly evaluate the fitness of each test case in isolation, it is needed to execute every test with the same state of SUT (i.e., test case executions should be independent from each other).
Thus, it requires to perform a state reset of the SUT before a test is executed on it, e.g., clear all data in the database or reset databases to a specific state.
With open-source case studies, it is trivial, e.g., clean data in database.
For instance,
\evo provides a utility \textit{DbClearner} for facilitating the cleaning of data for various types of SQL databases, e.g., Postgres and MySQL.
Such a clean on the database does work fine for small-scale applications.
However, in large-scale industrial settings, cleaning all data in the database is quite expensive, even when the database is empty.
For instance, in one of the industrial APIs used in this paper, it takes 5.3 seconds to clean an empty database, and it takes more time if there exist data.
Thus, within 1 hour as search budget, a fuzzer can execute at most 680 RPC function calls. 
This would significantly limit the fuzzer in terms of cost-effectiveness.
To better enable our approach in industrial settings, by taking advantage of existing \textit{SQL handling} in \evo, we developed an automated \emph{smart clean} on the database, by considering only what tables are actually modified during the search.
With the smart clean, after a test is executed, data only in the accessed tables and linked tables (e.g., with foreign key) will be removed.
In addition, we also allow SQL commands/scripts to initialize data into the database (e.g., for username/password authentication info).
If a table which has initial data is cleaned, a post action will be performed to add the initial data for the table again.
With such smart database clean, we could effectively reduce time spent by more than 90\%, e.g., from 5.3 seconds to 285 milliseconds.
This is because there can be tens/hundreds of tables in an industrial API, but only few of them are actually accessed during the executing of a single test. 
However, how to reset the state of the databases with a large amount of existing data still needs to be addressed.
Besides the database, the states of direct connected external services also need to be reset.
Currently, fuzzing by our approach is performed on the industrial test environment where all services are up and running.
With such an environment, the states of external services might be varied over time (e.g., failed tests as discussed in Section~\ref{subs:results-faults}).
Mocking technique could be a potential solution to address this, e.g., set up specific states of the external services before test execution.
However, mocking RPC-based services in microservices is also challenging, e.g., due to network communications and environment setup in industrial settings.
It could be considered as important future work.

\textbf{Real industrial APIs have more complex inputs and apply more restrict constraints in input validations with considerations of various aspects}.
By checking code coverage and fault detection, we found that most of codes and faults are related to the parts of implementation for input validation.
One reason could be due to the complexity of the input with cycle objects and collections in DTO.
For instance, we found that a DTO is initialized with more than 2\textit{k} lines, and generating a valid input for such a huge DTO would not be trivial.
In enterprise applications, often there exist several constraints on the inputs when processing their business logic.
This can  lead to major challenges for automated testing approaches to generate such inputs. 
The input validation is performed at two levels, i.e., in the schema and business logic.
The schema level would perform simple checks (e.g., null, format and range) and checks on constraints related to multiple fields in inputs.
Although we have supported the handling of all these constraints defined with \texttt{javax} annotations, it is clear that it is not enough in industrial settings.
Because not all constraints are  fully specified in the interface definitions, e.g., with \texttt{javax.validation.constraints}, the validation could be implemented as a utility or with libraries, e.g., \texttt{com.google.common.base.Preconditions}, directly in the code of the business logic. 
To address this, further white-box handling is required to provide more effective gradient to cover the code.

Regarding the validation in terms of business logic, it could perform a check with database and external services.
\textbf{Data preparation in database and mocking external services would be vital in testing of industrial RPC-based APIs}, not only for input validation.
For databases, currently our approach employs the \textit{SQL handling} in \evo~\cite{arcuri2020handling} for facilitating data preparation in the database.
However, as identified in this study, there might exist some limitations in handling industrial settings cost-effectively, e.g., currently \evo lacks support for composite primary keys.
This does limit the performance on code coverage.
For instance, we found that a query action with no input parameters is always failing with an exception thrown.
In this case, we could do nothing by manipulating the input parameters.
Then, with a manual check on the source code, we found that the query is required to have data in the database, but the data fails to be generated
 due to some unsupported SQL features.
In addition, with only SQL query heuristics, it might not be cost-effective to build meaningful links between RPC function calls and inserted data into the database. 
Smart strategies would be required here to handle industrial RPC-based APIs, such as the enhanced SQL handling strategies for REST APIs~\cite{zhang2021enhancing}. 
For external services, if we could mock such external services, then the problem might be solved by directly manipulating their responses. 
Automating such manipulation as parts of the search would be another important challenge.

Another possibility to improve code coverage would be to develop advanced search operators for the RPC domain.
For instance, we found that, in the generated tests, function calls in a test may not be related to each other for testing a meaningful scenario.
In order to better generate tests with related function calls, we could have strategies to sample function calls by considering dependency among functions (e.g., \cite{zhang2019resource}) in the context of RPC testing, e.g., based on which SQL tables they do access.

\textbf{An industrial RPC-based API is often a part of large-scale microservices that closely interacts with multiple APIs. Such interaction would result in a huge search space}.
To test a single API or an API in a small-scale microservice system, testing targets (such as lines of code) could be feasible to reach with an empty database (with/without a small amount of data initialized by SQL script) by manipulating input parameters and data into the database (e.g., \texttt{INSERT}).
However, testing an industrial API in microservices  is not like this case.
As the example shown in Figure~\ref{fig:rpc-context}, 
\textbf{the states of other services and databases often have a strong impact in processing business logic that would result in code coverage.}
Therefore, all such possible states are considered as a part of search.
In this paper, we provide descriptive statistics for {\totalcs} industrial APIs with \loc$_j$ (in total {\totalLoc}).
All of the APIs are parts of one microservice architecture, and there exist hundreds of other APIs which were not used in these experiments. 
To cope with such a huge complexity of the state,
an empty database (as we employed) might limit performance.
In addition, as discussed with our industrial partner, they think that\textbf{ it is important to involve their real historical data (collected in production) in the automated testing}.
Likely it would improve the chances to cover more of their business scenarios in the generated tests.
Furthermore, such tests would be more valuable for them, e.g., they would consider that all faults identified by these tests would have higher priority to be addressed.
However, such data is complex and possibly huge, and how to effectively and efficiently utilize this data with search would be another research challenge that we will address in the future. 

\textbf{Enabling fuzzers on CI would promote their adoption in industrial settings}.
Our approach  is now integrated into one of the industrial development pipelines (same as for the experiments we ran in this paper), as a trial to check its applicability into the daily testing activities of our industrial partner.
Since all services are developed with the same framework, by studying one of the \evo driver configurations for our approach, our industrial partner has implemented an automated solution to automatically generate such drivers for their services to be tested (e.g., identify all available interfaces and instantiate corresponding clients).
For instance, the drivers of the 30 industrial APIs in Table~\ref{tab:ind30} were automatically generated with this automated solution.
Regarding the application context, as discussed with our industrial partner, our approach is planned to be employed on the services for generating white-box system tests when the implementation for a requirement of the services is considered as done, as a kind of extra check before putting these new features in production.
In addition, the generated tests would be kept for further usage in (1) regression testing of the services and (2) industrial test environment validation as scheduled tasks (e.g., to see whether all services on the pipeline are up and running correctly before QA engineers start manual test sessions).

\textbf{Flakiness and readability are required to be considered in test generation in industrial APIs}.
As we found in the industrial APIs, responses could  contain info such as time-stamps and random tokens, and they could change over time.
In order to avoid test failing due to such flakiness, we defined some strategies with general keywords (e.g., date, token, time) observed in the industrial APIs to comment out assertions with such sources.
How to systematically identify possible sources of flakiness existing in the industrial APIs (e.g., time-stamps, results of SQL queries) and properly handle it during the automation and in the test generation would be another important problem that researchers should address.
During the process of reviewing the generated tests with industrial partner, 
we found that test readability requires to be improved.
This is mainly due to very large blocks of code  for input instantiations and large numbers of tests in the test suites.
As identified in the review, our industrial partner found that the tests which lead to exception thrown are more interesting for them.
Therefore, to improve test readability, we now provide a simple strategy to split such tests into different files (the implementation is straight-forwarded, but it is quite useful for our partner).
Further possible improvements could be achieved by better organizing the code for large input instantiations, and sorting/splitting tests based on various considerations, e.g., fault classification~\cite{marculescu2022faults}.

\section{Threats to Validity}
\label{sec:threats}

\emph{Conclusion validity}. 
Our study is in the context of SBST, and our experiments were conducted by following common guidelines in the literature to assess randomized techniques~\cite{Hitchhiker14}.
For instance, with a consideration for the stochastic nature of the employed search algorithms, we collected results for all settings with at least 30 repetitions.
The results were interpreted with statistical analysis, such as Mann-Whitney-Wilcoxon U-tests ($p$-$value$) and Vargha-Delaney effect sizes ($\hat{A}_{12}$) for pair comparisons.
Regarding fault detection capability, a number of real faults was identified, and those were reviewed together with our industrial partner.
Regarding the choice of search budget, since the time cost of executing RPC calls might vary depending on the operating environments (e.g., hardware and OS), we employed a fixed number of RPC calls as the search budget (i.e., 100 000), in order to make our experiment replicable.
Studying different settings of search budgets (such as 1 million RPC calls, 1 hour, 24 hours or 48 hours) might provide us more insights and more concrete evidence for drawing conclusions relating to the choice of the search budget (e.g., more budget might result in better performance on \csC and \csD, as discussed in Section~\ref{subsubsec:rq2_industrial_apis}).
However, it is expensive to conduct such an empirical experiment with industrial APIs, as these APIs are typically large-scaled and  complex.
For instance, with one search budget setting (i.e., 100 000 RPC calls), the computational cost of two settings with 30 repetitions is 129.12 days for the four APIs.
Therefore, we consider the experiments with various search budgets as possible future work.

\emph{Construct validity}. 
To avoid bias in the results among different settings and techniques, 
all results to be compared  were executed the on same physical environment, e.g., experiments on artificial case studies were deployed on a local machine, and experiments on industrial case studies were deployed on the pipeline of our industrial partners.

\emph{Internal validity}, 
Our implementation was tested with various units tests and end-2-end tests, but we cannot guarantee no fault in our implementation.
However, our tool and artificial case studies are open-source.
This enables further verification on our implementation and replication of our experiments on the artificial case studies by other researchers. 
Note that, due to the confidential info of the employed industrial case studies, detailed results of these industrial APIs cannot be made publicly available. 

\emph{External validity}. 
In this study, our approach was assessed with artificial case studies using Thrift and {\totalcs} industrial case studies (from one company) using their own RPC framework which is initially built based on Thrift.
There might exist a threat to generalize our results to other RPC frameworks or other companies.
Experiments on real-world industrial APIs show the usefulness and scalability of our novel techniques in practice.
However, these results cannot be replicated by other researchers, as such industrial APIs are not publicly available.
Collecting and preparing a corpus of non-trivial open-source RPC-based APIs for experimentation
(e.g., like EMB~\cite{EMB} for RESTful APIs) will be important for future research work.

\section{Conclusion}
\label{sec:conclusion}

RPC is widely applied in industry for developing large-scale distributed systems, such as microservices. 
However, automated testing of such systems is very challenging.
To the best of our knowledge, there does not exist any tool or solution in the research literature that could enable automated testing of such systems.
Therefore, having such a solution with tool support could bring significant benefits to industrial practice.

In this paper, we propose the first approach for automatically white-box fuzzing RPC-based APIs, using search-based techniques.
To access the RPC-based APIs, the approach is developed by extracting available RPC functions with \textit{RPCInterface}s from the source code.
This can enable its adoption to most RPC frameworks in the context of white-box testing.
To enable search techniques (e.g., MIO) in RPC domain, we reformulate the problem and propose additional handling and heuristics specialized for RPC.

The approach is implemented as an open-source tool built on top of our \evo~\cite{EvoMaster} fuzzer.
A detailed empirical study of our novel approach was conducted with two artificial and four industrial APIs, plus a preliminary (e.g., no fault analysis) study on a further {\totalindcs} APIs.
In total, more than a million lines of business code 
(excluding third-party libraries) 
were used in this study.
When third-party libraries are considered as well (e.g., for carrying out \emph{taint analysis}~\cite{arcuri2021tt}), several millions of lines of code were analyzed and executed in these experiments.

Results demonstrate the successful  applicability of our novel approach in industrial settings.
Our tool extension presented in this paper is already in daily use in the Continuous Integration systems of Meituan, a large e-commerce enterprise with hundreds of millions of customers. 
In addition, to evaluate the effectiveness of our approach in the context of white-box search-based testing, we compared our approach with a grey-box technique. 
The results show that our approach achieves significant improvements on code coverage. 
To further evaluate the capability of fault detection, we carried out an in-depth manual review with one employee of our industrial partner on the tests generated by our novel approach. 
A total of 41 real faults were identified that have now been fixed.
Another \totalIndfaults detected faults are currently under investigation.

Considering how widely used RPC frameworks such as Apache Thrift, Apache Dubbo, gRPC and SOFARPC have been in industry in the last decade, it can be surprising to see how such important software engineering topic has been practically ignored by the research community so far.
One possible explanation is the lack of easy access to case studies for researchers, as
these kinds of systems
are used to build enterprise applications.
Therefore, these systems are seldom available on open-source repositories, or online on the internet (i.e., general access web services are usually developed as REST APIs). 
To be able to empirically evaluate our novel techniques, industry collaborations (e.g., with Meituan) were a strong requirement.

Although our tool extension is already of use for practitioners in industry, more needs to be done to achieve better results.
Future work will focus on improving white-box heuristics to increase the achieved code coverage, and how to handle and analyze the interactions with external web services.

Our tool extension of \evo is freely available online on GitHub~\cite{EvoMaster} and Zenodo (e.g., \evo version 1.5.0~\cite{zenodo150evomaster}), and the replication package for this study can be found at the following link\textsuperscript{\ref{foot:link}}.

\section*{Acknowledgment}
This project has received funding from the European Research Council (ERC) under the European Union’s Horizon 2020 research and innovation programme (grant agreement No 864972).


\bibliographystyle{ACM-Reference-Format}

\end{document}